\documentclass[12pt]{iopart}
\usepackage{iopams}
\usepackage{epsfig} 
\usepackage{graphicx}
\def\lap{\mathrel{\mathpalette\fun <}}

\def\fun#1#2{\lower3.6pt\vbox{\baselineskip0pt\lineskip.9pt
\ialign{$\mathsurround=0pt#1\hfil##\hfil$\crcr#2\crcr\sim\crcr}}}
\def\lap{\mathrel{\mathpalette\fun <}}

\def\fun#1#2{\lower3.6pt\vbox{\baselineskip0pt\lineskip.9pt
  \ialign{$\mathsurround=0pt#1\hfil##\hfil$\crcr#2\crcr\sim\crcr}}}
\def\lap{\mathrel{\mathpalette\fun <}}

\newcommand{\MUNCH}[1]{\relax}

\begin{document}

\title{Prospects in Constraining the Dark Energy Potential}
\author{Enrique Fernandez-Martinez$^{1}$ \&  Licia Verde$^{2}$\\ 
$^1${\it  Max Planck Institut f\"ur Physik, F\"ohringer Ring, 6. M\"unchen, D-80805 Germany}  \\ 
$^2${\it  ICREA \& Institute of Space Sciences (IEEC-CSIC),  Fac. Ciencies, Campus UAB, Torre C5 parell 2, Bellaterra, Spain and Dept. of Astrophysical sciences, Peyton Hal, Ivy Lane, Princeton  
University, Princeton, NJ\\}}

\date{\today}

\begin{abstract} 
  We generalize  to non-flat geometries the formalism of Simon et al. (2005) to reconstruct the dark energy potential. This formalism  makes use of quantities similar
  to the Horizon-flow parameters in inflation, can, in principle,  be made non-parametric and is general enough to be applied outside the simple, single scalar field quintessence.
  Since presently available and forthcoming data do not allow a
  non-parametric and exact reconstruction of the potential, we
  consider a general parametric description in term of Chebyshev polynomials. We then consider present and future measurements of $H(z)$, Baryon Acoustic Oscillations surveys and Supernovae type 1A surveys, and investigate their constraints on the dark energy potential.
We find that, relaxing the flatness assumption increases the errors on the reconstructed dark energy evolution but does not open up significant degeneracies, provided that a modest prior on geometry is imposed.   Direct  measurements of $H(z)$, such as those provided by BAO surveys,  are crucially important to constrain the evolution of the dark energy potential and the dark energy equation of state, especially for non-trivial deviations from the standard $\Lambda$CDM model.    

\end{abstract}

%\pacs{} 

\maketitle

\section{Introduction}
\label{sec:intro}
Recent observations e.g., \cite{SpergelWMAP03,SpergelWMAP06,wmap5dunkley,wmap5komatsu,  Wood-VaseySN07, TegmarkLRGDR4,PercivalLRG} indicate  that the  present-day energy density of the universe is dominated  by a ``dark energy" component, responsible for the current accelerated expansion.

The leading  dark energy candidates are a cosmological constant or a slowly varying rolling scalar field  e.g.,\cite{Caldwell98, Quint99,Wang00,RatraPeebles88a,RatraPeebles88b,RatraPeebles95,Wetterich95} and \cite{Steinhardt03} for a review, although explanations in terms of modifications of the Friedman equation (\cite{DGP,DDG02,Carroll03,Carroll05,Capozziello05,NojiriOdintsov04,Dolgov03,HuSaw07, Acquaviva05,AV07} and references therein, under the name of ``modifications of gravity" and e.g.,\cite{Enqvist08, rasanen06,KKNNY07,KolbMatarreseRiotto06}  under the name of inhomogeneous models) are also being widely investigated.

An extended observational effort is being carried out  (e.g., SNLS, SDSS,  PanStarrs etc.)  and ambitious plans for the future are being proposed or planned (e.g., DES, PAU, BOSS, WFMOS, SNAP,  ADEPT,  DUNE, SPACE, SKA, LSST) with the goal of shedding light on the nature of dark energy.
 With few exceptions, current constraints on the nature of dark energy  measure an integrated value over time of the Hubble parameter, $H(z)$, which in turn is an integral of  its equation of state parameter ($w=p/\rho$, with $p$ denoting pressure  and $\rho$ density).  While these constraints are very tightly centered around the cosmological constant value, with a ~15\% error, the finding that the average value of $w$ is consistent with $-1$ does not exclude the possibility that $w$ varied in time e.g., \cite{Maor02}. 
 An emerging technique in dark energy studies uses observations of the so-called baryon acoustic oscillations (BAO) \cite{EH98, Eisenstein05, 2df05}.  The BAO  yield a measurement of the sound horizon at recombination,  a standard ruler visible at  different epochs in the lifetime of the Universe: at the last scattering surface through cosmic microwave background observations and  at lower redshifts  through galaxy clustering.
 In galaxy surveys,  the BAO scale  can be measured both along and perpendicular to the line sight. In particular the line-of-sight measurement offers the unique opportunity to measure directly $H(z)$,  rather than  its integral.
 
To improve our understanding of dark energy, it is important   not only to ask whether this dark energy component is dynamical or constant, but also,  to constrain possible shapes of the dark energy potential. As different theoretical models are characterized by different potentials, a reconstruction of the dark energy potential  from cosmological observations could help discriminating among different, physically motivated, models.

Different approaches to the reconstruction of the dark energy equation of  
state or potential have been proposed in the literature e.g., \cite {S1,HutererTurner01,HS03, Simonetal05,S2,HP07}. In this paper we build upon and generalize the reconstruction technique proposed in \cite{Simonetal05}  to non-flat universes.
In fact \cite{Polarski:2005jr,ClarksonCortesBassett07} showed that there can be a degeneracy between geometry and dark energy properties, thus  analyses to constrain dark energy parameters  should be carried out  varying  jointly the geometry of the Universe. 

We then apply  this reconstruction  to  existing determinations of $H(z)$  from ages of passively evolving galaxies \cite{Simonetal05}, to new  Supernovae data \cite{Wood-VaseySN07}, and  we forecast the constraints on the dark energy potential achievable with  the next generation of BAO  and Supernova  surveys.  
We find that relaxing the assumption of flatness increases the error in the reconstructed dark energy evolution but does not open up degeneracies, provided that a modest prior on geometry is imposed: a Gaussian prior on  $\Omega_k$ with  {\it r.m.s.} $\sigma_k=0.03$.
Measurements of $H(z)$ such as those provided by BAO surveys, are crucial to constrain the dark energy evolution and to break degeneracies among dark energy parameters, for non-trivial deviations from the simplest $\Lambda$CDM model. 

The rest of the paper is organized as follows: in \S 2  we present our reconstruction of the dark energy potential.  In \S 3 we describe the priors  used in addition to the  present and future  data sets we consider. Our results on the dark energy potential reconstruction  are reported in \S 4.  For completeness, we present a similar reconstruction  applied to the dark energy equation of state parameter as a function of redshift (in \S 5). We conclude in \S 6.

\section{How to reconstruct the dark energy potential from observations}
\label{sec:procedure}
Ref.~\cite{Simonetal05} presented a non-parametric method to reconstruct the redshift evolution of the potential and kinetic energy densities of the dark energy field, using quantities similar to the Horizon-flow parameters in inflation  e.g., \cite{schwarz01, Leach02}. As a fully non-parametric reconstruction would require the knowledge of $H(z)$ {\em and} of $\dot{H}(z)$, \cite{Simonetal05} also  presented a general parameterization of the dark energy potential as a function of redshift, $V(z)$, in term of Chebyshev polynomials and  showed how to reconstruct  the potential as a function of the scalar field $V(\phi)$ from $V(z)$.  
In this section we will follow Ref.~\cite{Simonetal05} to directly relate the dark energy potential $V(\phi)$ with observable quantities, but we generalize their approach to non-flat universes. 

We  restrict ourselves to classical configurations $\phi=\phi(t)$, that do not break the homogeneity and isotropy of space-time.
The energy-momentum tensor of this scalar field configuration is that of a
perfect fluid, with density $\rho_\phi$ and pressure $p_\phi$ given by
\begin{equation}
  \rho_\phi = K(\phi)+ V(\phi)\,, \quad
  p_\phi = K(\phi)- V(\phi)\, \quad \mbox{and} \,\,\,\, K\equiv\frac{1}{2}\dot{\phi}^2 .
 \label{eq:qtensor}
\end{equation}
where $K$ denotes the kinetic energy of the field.
The Friedmann's equations then read:
\begin{eqnarray}
    H^2 & = \frac{\kappa}{3}\,\left(\rho_T + \rho_\phi + \rho_k\right)\,, \\
    \frac{\ddot{a}}{a} &= \frac{1}{2H}\frac{dH^2}{dt}= - \frac{\kappa}{6}\left(\rho_T + 3p_T 
    + \rho_\phi + 3\,p_\phi\right)~,
 \label{eq:friedmann}
\end{eqnarray}
where $\kappa = 8\pi/m_p^2$ (or $\kappa = 8\pi G$) and $\rho_k = -k\frac{3c^3}{\kappa a^2}$, where $k$ is the curvature. In Eq. (\ref{eq:friedmann})  we introduced
the compact notation $\rho_T$ and $p_T$ for the total energy density
and pressure. Using both Friedmann's equations to solve for the kinetic energy of the field and considering a single matter component we obtain:
\begin{equation}
  3H^2(z) - \frac{1}{2}\,(1+z)\,\frac{d\,H^2(z)}{dz} =
  \kappa\,\left(V(\,z) + \frac{1}{2}\rho_m(z) + \frac{2}{3}\rho_k(z)\right)~,
 \label{eq:Hz}
\end{equation}
An exact non-parametric reconstruction
of $V(z)$ is possible only if $H(z)$ and $\dot{H}(z)$ are known (see Ref.~\cite{Simonetal05} for details). 
Unfortunately present and near future prospective data do not allow this level of precision.  However if $V(z)$ can be parameterized, Eq. (\ref{eq:Hz})
 can be integrated analytically:
\begin{eqnarray}
  H^2(z) &=&  \left(H_0^2-\frac{\kappa}{3} (\rho_{m,0}+\rho_{k,0})\right)\,(1+z)^6  \nonumber \\
   &+&
  \frac{\kappa}{3} (\rho_m(z)+\rho_k(z))
  -2(1+z)^6\int_0^z\,V(x)\,(1+x)^{-7}\,dx~.
 \label{eq:Hzsol}
\end{eqnarray}
Hereafter the $0$ subscript denotes the quantity evaluated at $z=0$. 

 An interesting  parameterization of the potential involves the
Chebyshev polynomials, which form a complete set of orthonormal functions
on the interval $[-1,1]$. They also have the interesting property to be 
the minimax approximating polynomial, that is, the approximating polynomial 
which has the smallest maximum deviation from the true function at any given 
order. We can thus approximate a generic $V(z)$ as
\begin{equation}
V(z)\simeq\sum_{n=0}^{N}\lambda_n T_n(x)
\label{eq:V-cheb-expan}
\end{equation}
where $T_n$ denotes the Chebyshev polynomial of order $n$ and  we have normalized the redshift interval so that
$x=2z/z_{max}-1$; $z_{max}$ is the maximum redshift at which
observations are available and thus $x\in [-1,1]$. Since $|T_n(x)| \le
1$ for all $n$, for most applications, an estimate of the error
introduced by this approximation is given by $\lambda_{N+1}$.  With
this parameterization, the relevant integral in Eq. (\ref{eq:Hzsol})
becomes:
\begin{eqnarray}
  \int_0^zV(y)(1+y)^{-7}dy &=&\frac{z_{max}}{2}\sum_{n=0}^N\lambda_n\int_{-1}^{2z/z_{max}-1}
  T_n(x)(a+bx)^{-7}dx \nonumber \\
  &\equiv&  \frac{z_{max}}{2} \sum_{n=0}^{N}\lambda_n\,F_n(z)
 \label{eq:chebxzpn}
\end{eqnarray}
where $a=1+z_{max}/2$ and $b=z_{max}/2$.
These integrals can be solved analytically for any order $n$ and 
$F_n$ are known analytic functions. Substituting in (\ref{eq:Hzsol}) we finally obtain:
\begin{eqnarray}
H^2(z,\lambda_i)=(1+z)^6 H_0^2\left[1 -3 z_{max} \sum_{n=0}^N
  \frac{\lambda_n}{\rho_{c}} F_n(z)\right. \nonumber \\
  \left.-\Omega_{m,0}\left(1-\frac{1}{(1+z)^3} \right)
-\Omega_{k,0}\left(1-\frac{1}{(1+z)^4} \right)\right]~,
\label{eq:Hsq-cheby-curv}
\end{eqnarray}

which relates observable quantities such as the Hubble parameter and $\Omega_{m,0}$ with the coefficients of the Chebyshev expansion of the potential.
Determining the coefficients $\lambda_i$ in this manner allows one to reconstruct $V(z)$ as in (\ref{eq:V-cheb-expan}). To obtain $V(\phi)$, however, we would
also need to reconstruct $\phi(z)$. This can also be accomplished from the determination of the coefficients $\lambda_i$, through the kinetic energy of the field. 
From the first Friedmann equation we have:
\begin{eqnarray}
  K(z)=\frac{1}{2}\left(\frac{d\phi}{dz}\right)^2\,(1+z)^2\,H^2(\lambda_i,\,z)\nonumber \\
  = 3\kappa^{-1}\,H^2(\lambda_i,\,z) - \rho_m(z) - \rho_k(z) - V(\lambda_i,\,z)~,
 \label{eq:qz}
\end{eqnarray}
which can be integrated to obtain $\phi(z)$ and thus $V(\lambda_i,\,\phi)$
from $V(\alpha_i,\,z)$:
\begin{eqnarray}
  \phi(z)-\phi(0) = \nonumber \\
  \pm \int^z_0\,
  \frac{\sqrt{6\kappa^{-1}\,
  H^2(\lambda_i,\,z) - 2\rho_m(z) - 2\rho_k(z) - 2V(\lambda_i,\,z)}}{(1+z)\,H(\lambda_i,\,z)}\,dz \label{eq:qzsol}
\end{eqnarray}
where the ambiguity in sign comes from the quadratic expression for the kinetic
energy. Typically, if we think of a scalar field rolling slowly along its potential, the plus sign will be the relevant one.

In what follows we will only consider a three parameters model (i.e. $N=2$). Even with only three parameters, if the fiducial model is a $\Lambda$CDM, the  forecasted constraints on deviations from a flat potential are rather weak especially at $z>0.5$. However, with the formulation presented here, one can pose the question of how many dark energy parameters are required by the data. Techniques such as cross validation \cite{GreenSilverman,sealfon05,verdePeiris08} could be used to address the issue.

In the next sections we describe the different observables we consider to probe the dark energy potential.

\section{Datasets and priors}
Most probes of dark energy measure integrated quantities, for example Supernovae measure  the luminosity distance $d_L(z)$.
There are two known  techniques to reconstruct directly $H(z)$. One is  through  the measurement of the BAO scale in the radial direction through $dr=c/H(z)dz$. Current surveys do not yet have the  sufficient statistical  power to do so and thus current measurements are angle-averaged. However, forthcoming and future surveys promise to deliver $H(z)$ determination with \% accuracy. BAO surveys, of course, also provide measurement of the angular diameter distance $d_A(z)$.

The other technique relies on  the measurement of ages of passively evolving galaxies. It has been demonstrated by recent observations that massive ($L > 2 L_*$) luminous red galaxies have formed more than 95\% of their stars at redshifts $> 4$. Since then, stars in these galaxies have been evolving passively. They are therefore, excellent cosmic clocks, where the age of their stars can be inferred from the integrated stellar spectrum using stellar evolution theory.

BAO surveys along with large samples of type 1A Supernovae are among the leading techniques to constrain dark energy and an extensive experimental effort is being carried out. For this reason we  consider presently available ``cosmic clocks" data, presently available Supernovae data and  future  BAO and  Type 1A supernova surveys.

\subsection{Priors}

In what follows we assume a flat $\Lambda$CDM with $H_0= 73.2$ $\textrm{Km s}^{-1} \textrm{Mpc}^{-1}$ and $\Omega_m = 0.24$ fiducial model.  In all cases we consider Gaussian priors of  $\sigma_H=8$ $\textrm{Km s}^{-1} \textrm{Mpc}^{-1}$ for $H_0$, $\sigma_{w_m}=0.01$ for $\Omega_m h^2$ and $\sigma_k=0.03$ for $\Omega_k$.  This is motivated by the fact that current data already constrain these parameters at this level e.g., \cite{Freedman2001,SpergelWMAP06, wmap5komatsu}. We will assume here that  these parameters can be constrained at this level  by  combination of e.g., Cosmic Microwave Background experiments  (e.g., Planck \cite{Planck}) and local determinations of the Hubble parameter \cite{riess2009inprep}.

\subsection{Baryon Acoustic Oscillations}
\label{sec:BAO}

Dark matter overdensities in the early Universe produce acoustic waves in the photon-baryon plasma that propagate with the speed of sound until the recombination era, 
when photons decouple from baryons and free stream. The baryon wave then stops propagating leaving an imprint at a characteristic distance from the original dark matter overdensity: the sound horizon length. This process thus provides a standard ruler at which the correlation function of dark matter (and thus of galaxies)  should peak (e.g., \cite{EH98}). 
Evidence of this peak has already been reported in galaxy surveys e.g., \cite{Eisenstein05,2df05}.
Measuring this standard ruler at different redshifts would provide a powerful probe of the expansion history of the Universe and thus of the dark energy potential.

Baryon Acoustic Oscillations (BAO) can be measured both along and perpendicular to the line sight. An angular measurement of the BAO scale at  redshift $z$ 
would then give:
\begin{equation}
\Delta \theta = \frac{r_{BAO}}{(1+z)d_A(z)}~,
\end{equation}

\noindent
where $d_A(z) = 1/(1+z)\int_{0}^{z} c/H(z) dz$ and $r_{BAO}$ is the BAO scale. This measurements of $d_A(z)$ can then be compared to Eqs. (\ref{eq:Hsq-cheby-curv}) or (\ref{eq:Hzwcheby}, below in \S 5.) to derive constraints on the coefficients $\lambda_i$ or $w_i$ respectively. Alternatively, if the redshift precision of the survey is good enough,
the BAO scale could be measured along the line of sight as 
\begin{equation}
\Delta z = H(z) r_{BAO}~,
\end{equation}
thus providing a direct measurement of $H(z)$.

To forecast the errors with which $H(z)$ and $d_A(z)$ will be recovered we make use of the formulas derived in \cite{Blake:2005jd}, where a grid of
BAO surveys was simulated with different survey parameters and the accuracy found for the observables was fitted to the following formulas:
\begin{equation}
\sigma_d(z_i)=x_0^d \frac{4}{3} \sqrt{\frac{V_0}{V_i}}f_{nl}(z_i)
\label{errord}
\end{equation}
\begin{equation}
\sigma_H(z_i)=x_0^H \frac{4}{3} \sqrt{\frac{V_0}{V_i}}f_{nl}(z_i)
\label{errorH}
\end{equation}
where
\begin{equation}
f_{nl}(z_i)= \left\{
\begin{array}{cc}
1 & z<z_m\\
(\frac{z_m}{z_i})^{\gamma} & z>z_m
\end{array}
\right.
\label{fnl}
\end{equation}
$V_i$ is the volume of the redshift bin $z_i$. The fitting formula was motivated by the assumption 
that the accuracy achievable in the observables will be proportional to the fractional error with which the
power spectrum can be recovered
\begin{equation}
\frac{\Delta_P}{P} \simeq \sqrt{\frac{2}{N_m}} \left( 1 + \frac{1}{nP} \right)~,
\label{errorP}
\end{equation}
where $N_m \propto V_i$ is the number of of Fourier modes contributing to the measurement and $n$ the number density of galaxies surveyed. Non-linearities tend to erase the acoustic peaks via mode-coupling: the function in (\ref{fnl}) 
takes into account that, at increasing redshift, increasingly small scales are in  the linear regime. The fitting parameters where  calibrated on N-body simulations by \cite{Blake:2005jd} and found to be:
$x_0^H=0.0148$; $x_0^d=0.0085$; $V_0=2.16/h^3$; $z_m=1.4$; $\gamma=0.5$.

Here we will consider two setups that  roughly encompass ground-based (``ground") and space-based (``space") perspective BAO surveys. The survey parameters are summarized in Table \ref{tab:survey}. In both cases we assume that shot noise is unimportant at the scale of interest and that the redshift determination is good enough to measure  the radial  BAO signal. Along the way we will also report the results for the angular-only BAO. This case will be relevant to photometric surveys that can achieve photometric errors better than $\sim 4$\% \cite{SeoEisenstein03, BlakeBridle05}

\begin{table}[hbtp]
\begin{center}
\begin{tabular}{|c|c|c|c|c|} 
\hline
Survey & Area $(dg^2)$ & $z_{min}$ & $z_{max}$ & bins in $z$ \\ 
\hline
\hline
ground & 10000 & 0.1  & 1 & 9 \\
\hline
space & 30000 & 1  & 2 & 10 \\
\hline
\end{tabular}
\end{center}
\caption{\it 
Survey parameters of the two BAO surveys considered.}
\label{tab:survey}
\end{table}

\subsection{Galaxy Ages}
\label{sec:ages}
The Hubble parameter depends on the differential age of the Universe as a function of redshift via
$H(z)=dz/dt (1+z)^{-1}$. The feasibility of measuring $H(z)$  from high-resolution, high signal-to-noise spectra of passively evolving galaxies was demonstrated in \cite{ages03, Simonetal05, ages08}. Here we use the $H(z)$ determination obtained by \cite{Simonetal05} and publicly available at \cite{webpage},  from a compilation of data at $0<z<1.8$ and  generalize the analysis of  \cite{Simonetal05} to non flat universes.
Recent studies \cite{Treu05,Cooletal08}  have clearly established that massive ($>2.2L_*$ ) luminous red galaxies have formed more  than 95\% of their stars at redshifts higher than 4. These galaxies, therefore, form a very 
uniform population, whose stars are evolving passively after the very first short episode 
of star active star formation \cite{Treu05,Cooletal08,Heavens04}. Because the stars evolve passively, these massive 
LRG are excellent cosmic clocks, i.e. they provide a direct measurement of $dt/dz$; the 
observational evidence discards further star formation activity in these galaxies. Dating 
of the stellar population can be achieved by modeling the integrated light of the stellar 
population using synthetic stellar population models, in a similar way to what is done 
for open and globular clusters in the Milky Way. The dating of the stellar population 
needs to be done on the integrated spectrum because individual stars are not resolved 
and therefore the requirements on the observed spectrum are stringent as one needs a 
very wide wavelength coverage, spectral resolution and very high signal-to-noise. Ref. \cite{Cooletal08} has shown that the spectra of these massive LRG at a redshift  $\sim  0.15$ 
are extremely similar, with differences of only 0.20 mmag, which is another evidence 
of the uniformity of the stellar populations in these galaxies. There have been already 
examples of accurate dating of the stellar populations in LRGs (\cite{Dunlopetal96, 
Spinradetal97,  ages03, Simonetal05}) where it has been  
shown that galaxy spectra with sufficient wavelength coverage (the UV region is crucial), 
wavelength resolution (about 3 \AA) and enough S/N (at least 20 per resolution element 
of 3 \AA) can provide sensible constraints on cosmological parameters. More details can 
be found in \cite{Simonetal05,Sternetalinprep}. 

\subsection{Supernovae}
\label{sec:SN}

The intrinsic luminosity of Type IA Supernovae (SN) can be accurately predicted from the decay rate of the Supernovae brightness. This provides bright standard candles that can be observed up to redshifts $z>1$. Measuring SN apparent magnitude, the luminosity distance can thus be inferred:  
\begin{equation}
d_L(z) = (1+z)\int_{0}^{z} \frac{c}{H(z)} dz~.
\end{equation}
If their redshift is also measured, Type IA Supernova provide information on the integral of $1/H(z)$ and hence on the cosmological parameters. In this way Supernovae provided
the first  direct evidence for the accelerated expansion of the Universe \cite{Riess:1998cb, Perlmutteretal1998}.

We consider present and forecasted Type IA Supernovae data in the analysis of Section \ref{sec:results}. For the present Supernovae data we use the sample of \cite{Davis07}.
For future Supernovae data we assume 1000 Supernovae distributed in 5 redshift bins between 0.8 and 1.3 plus a sample of 500 Supernovae at low redshift \cite{DETF, AV07}. Table \ref{tab:sn} summarizes the distribution of the Supernovae considered.

\begin{table}[hbtp]
\begin{center}
\begin{tabular}{|c|c|c|c|c|c|c|} 
\hline
Mean z & 0.1 & 0.85 & 0.95 & 1.05 & 1.15 & 1.25 \\ 
\hline
SN & 500 & 231 & 219 & 200 & 183 & 167 \\
\hline
\end{tabular}
\end{center}
\caption{\it 
Redshift distribution of the forecasted Supernovae sample.}
\label{tab:sn}
\end{table}

We consider a statistical error  on $\mu = 5 \log d_L + K$  of $\sigma_{\mu, stat.}= 0.1$ due to the uncertainty of the corrected apparent magnitudes. We also consider a systematic error given by \cite{Linder:2002bb} $\sigma_{\mu, syst.}=0.02(1+z)/2.7$.

For both Supernovae samples, present and forecasted, we marginalize over the absolute magnitude of the sample.

\section{Results}
\label{sec:results}

For the different  data sets  we  compute (or  forecast, for future data)  the constraints on the first three coefficients of the Chebyshev expansion, assuming a flat $\Lambda$CDM with $H_0= 73.2$ $\textrm{Km s}^{-1} \textrm{Mpc}^{-1}$ and $\Omega_m = 0.24$ fiducial model with the priors described above and errors on $H(z)$ and $d_A(z)$  and $d_L(z)$ as outlined in \S 3. For Supernovae and galaxy ages data the analysis is performed exploring the likelihood surface via a Markov Chain Monte Carlo.
For the  BAO surveys   we  use both  a  Markov Chain Monte Carlo  and a Fisher matrix approach, finding good agreement between the two techniques. Here we present the results of the Fisher matrix analysis. 

\begin{figure}[t]
\vspace{-0.5cm}
\begin{center}
\begin{tabular}{cc}
\hspace{-0.55cm} \includegraphics[width=7.5cm]{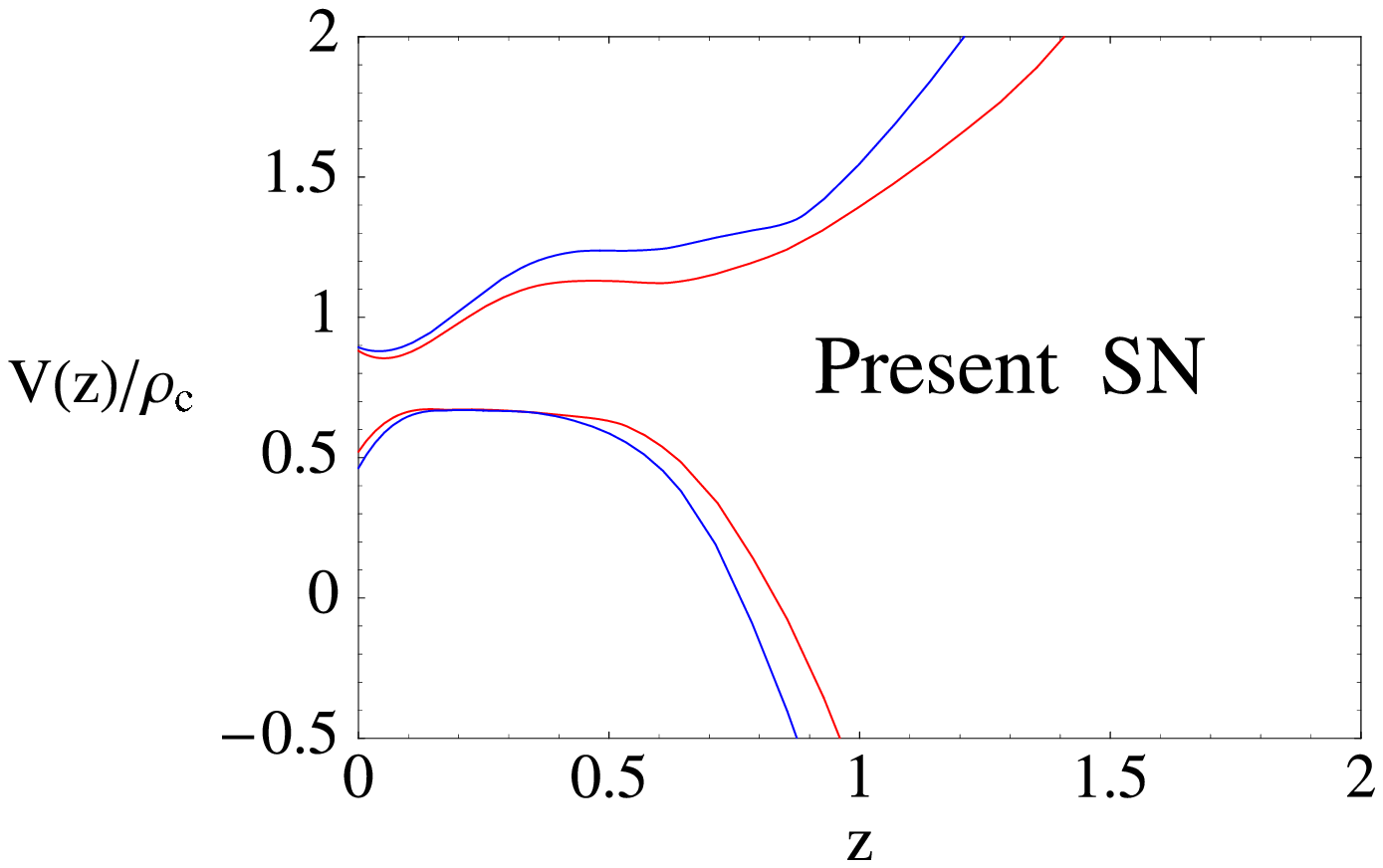} &
	 \includegraphics[width=7.5cm]{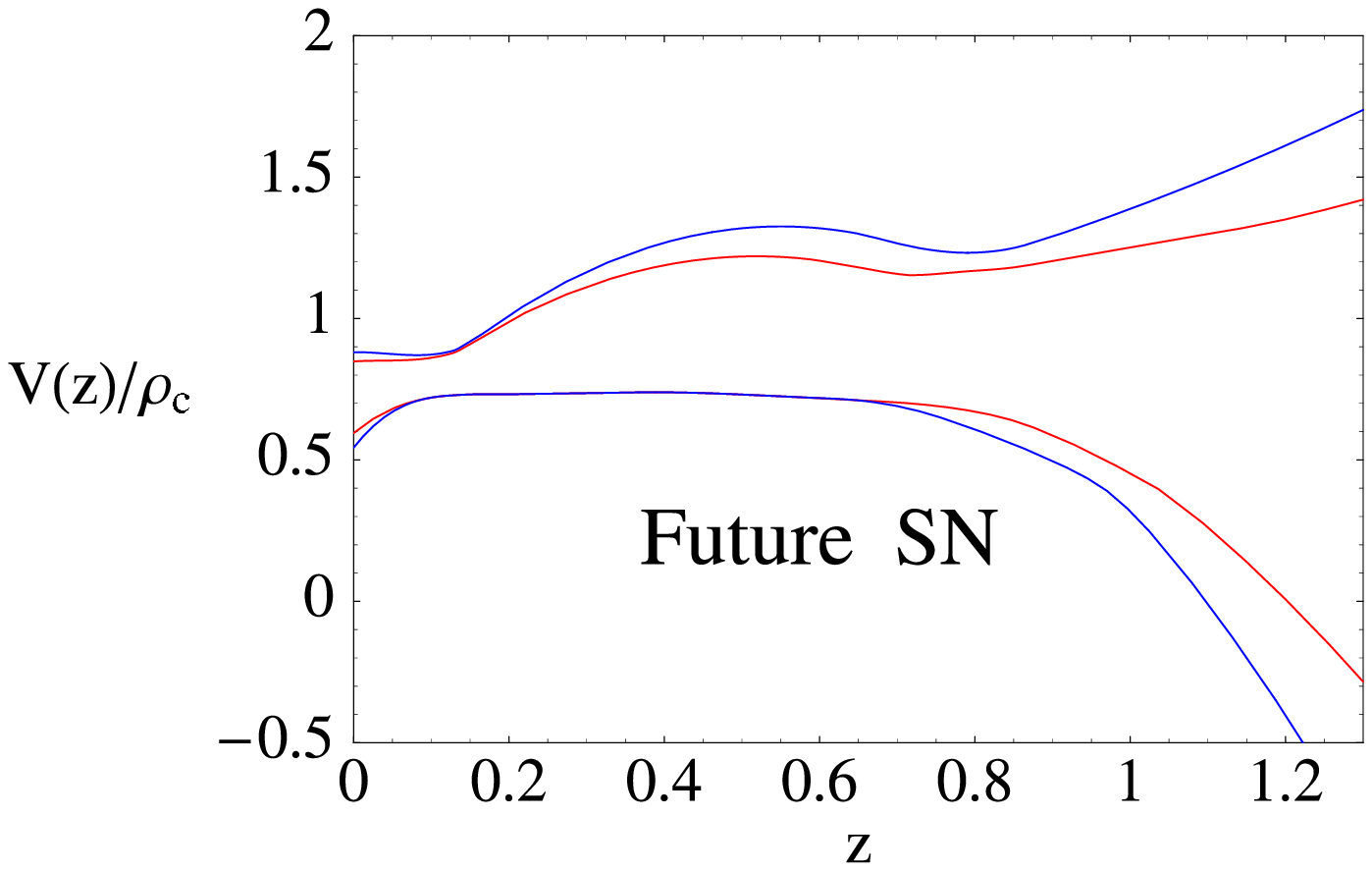} \\
\hspace{-0.55cm} \includegraphics[width=7.5cm]{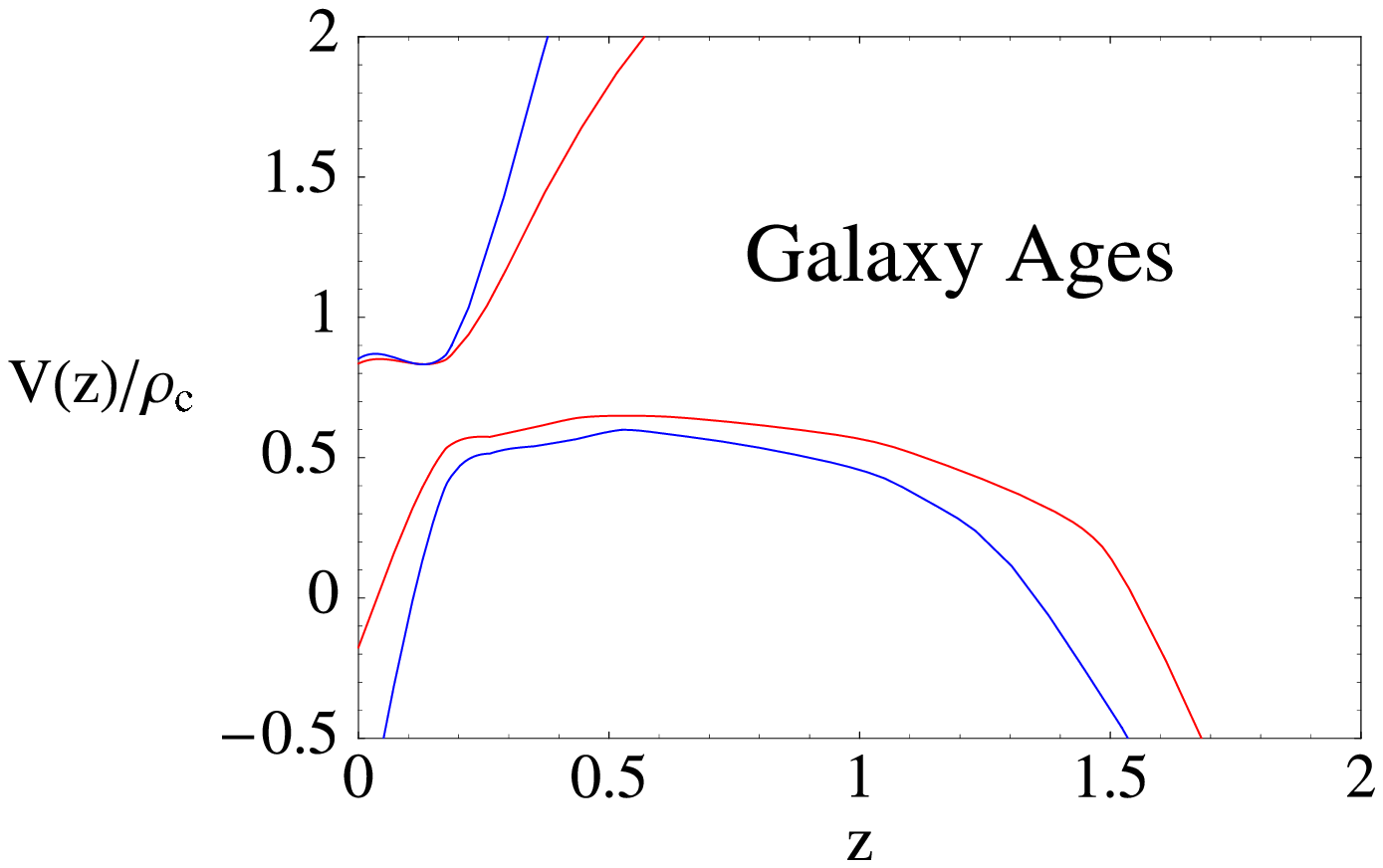} &
	 \includegraphics[width=7.5cm]{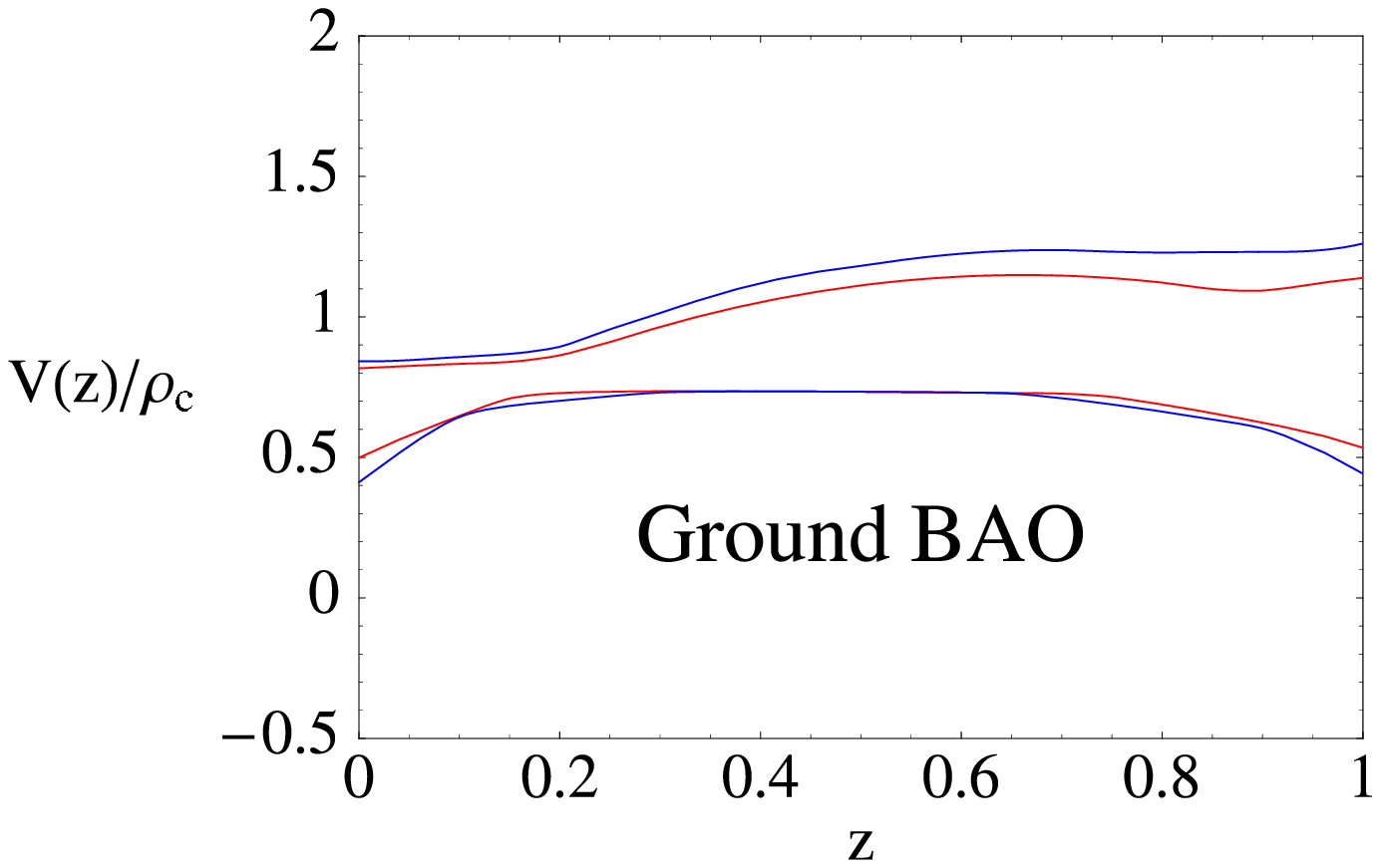} \\
\hspace{-0.55cm} \includegraphics[width=7.5cm]{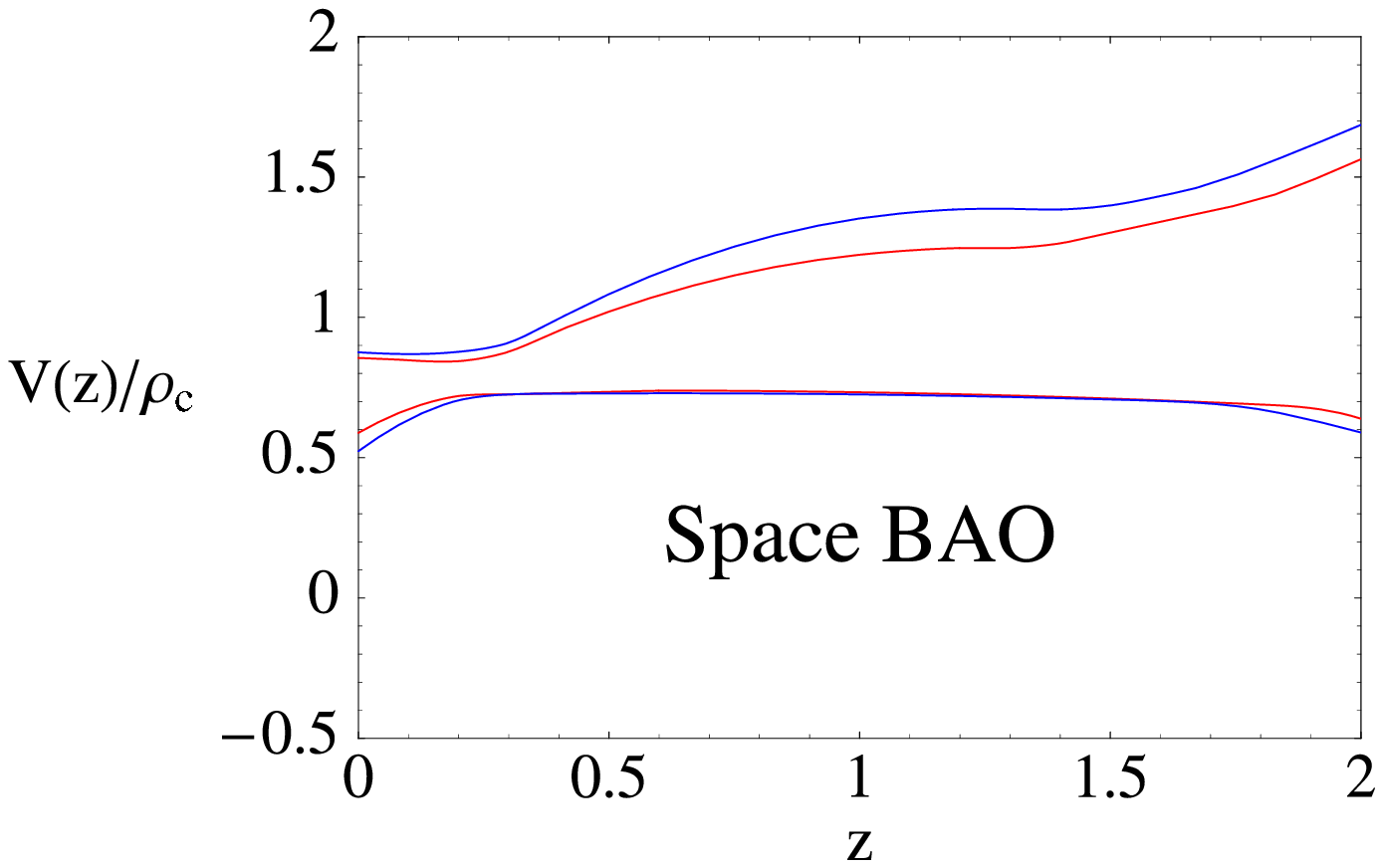} &
		 \includegraphics[width=7.5cm]{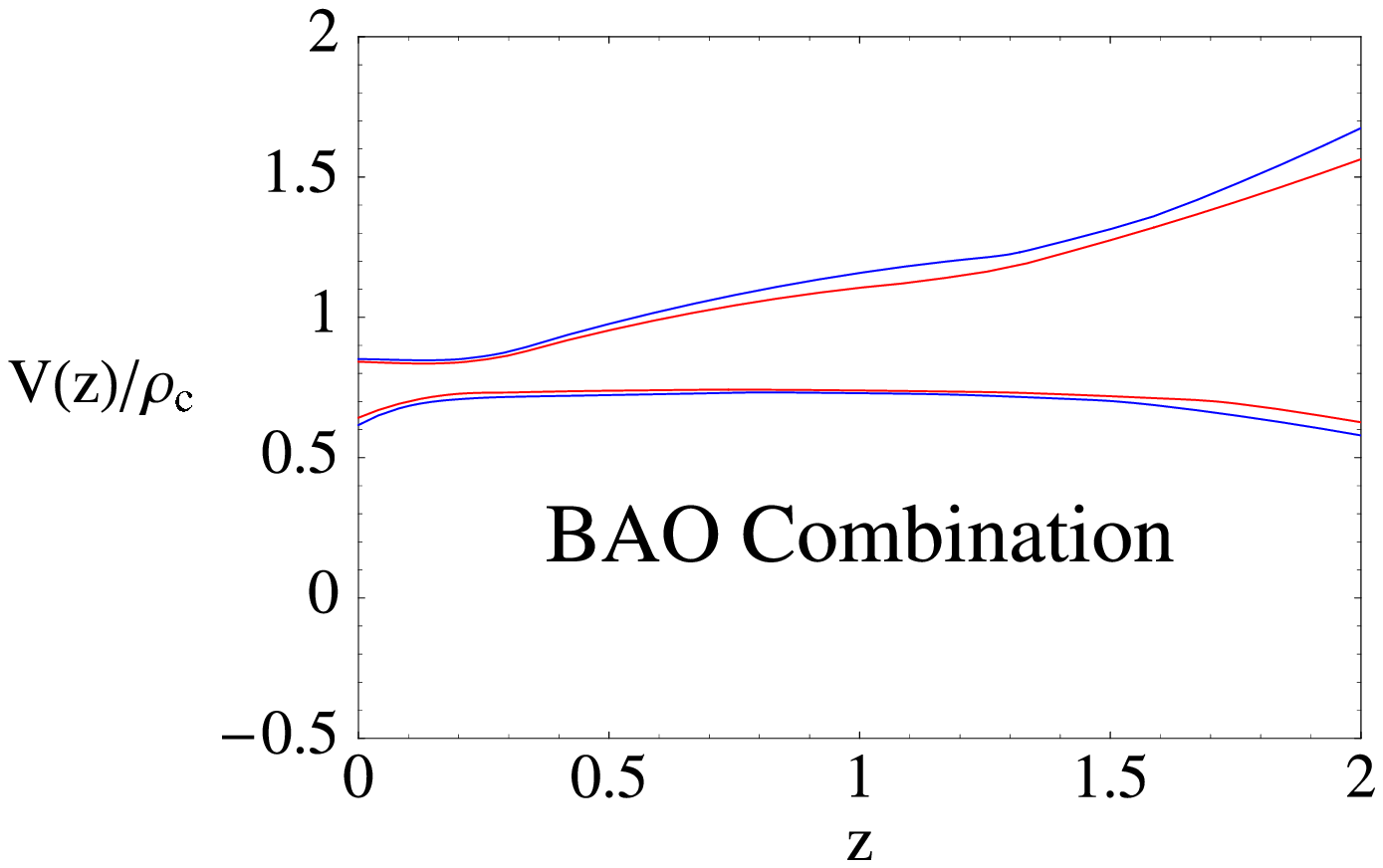}
\end{tabular}
\caption{\it 1 and 2 $\sigma$ constraints on the reconstructed  potential as a function of redshift $V(z)$ from present SN data (top left), SN data from a future-space-based experiment (top right), galaxy ages (middle left), ``ground" BAO survey (middle right), ``space" BAO survey  (bottom left) and the combination of the two BAO surveys (bottom right).}
\label{fig:potential}
\end{center}
\end{figure}

The constraints on the $\lambda_i$ thus derived can be translated into constraints on the potential using Eq.~(\ref{eq:V-cheb-expan}). In Fig.~\ref{fig:potential} we show
the results of this reconstruction for Supernovae (present and future, top panels) galaxy ages (left middle panel),  ``ground" BAO survey (middle right), ``space" BAO survey (lower left) and the combination of the two BAO surveys (lower right).

Notice that the constraints set by all the datasets considered are strongest between $z \sim 0.1-0.3$. This is a consequence of the recent dominance of dark energy in the cosmological history and translates in a strong linear degeneracy between $\lambda_0$ and $\lambda_1$  as discussed below.

\subsection{Interpretation of the reconstructed $V(z)$}

\begin{figure}[t]
\vspace{-0cm}
\begin{center}
\begin{tabular}{cc}
\hspace{-0.55cm} \includegraphics[width=7.5cm]{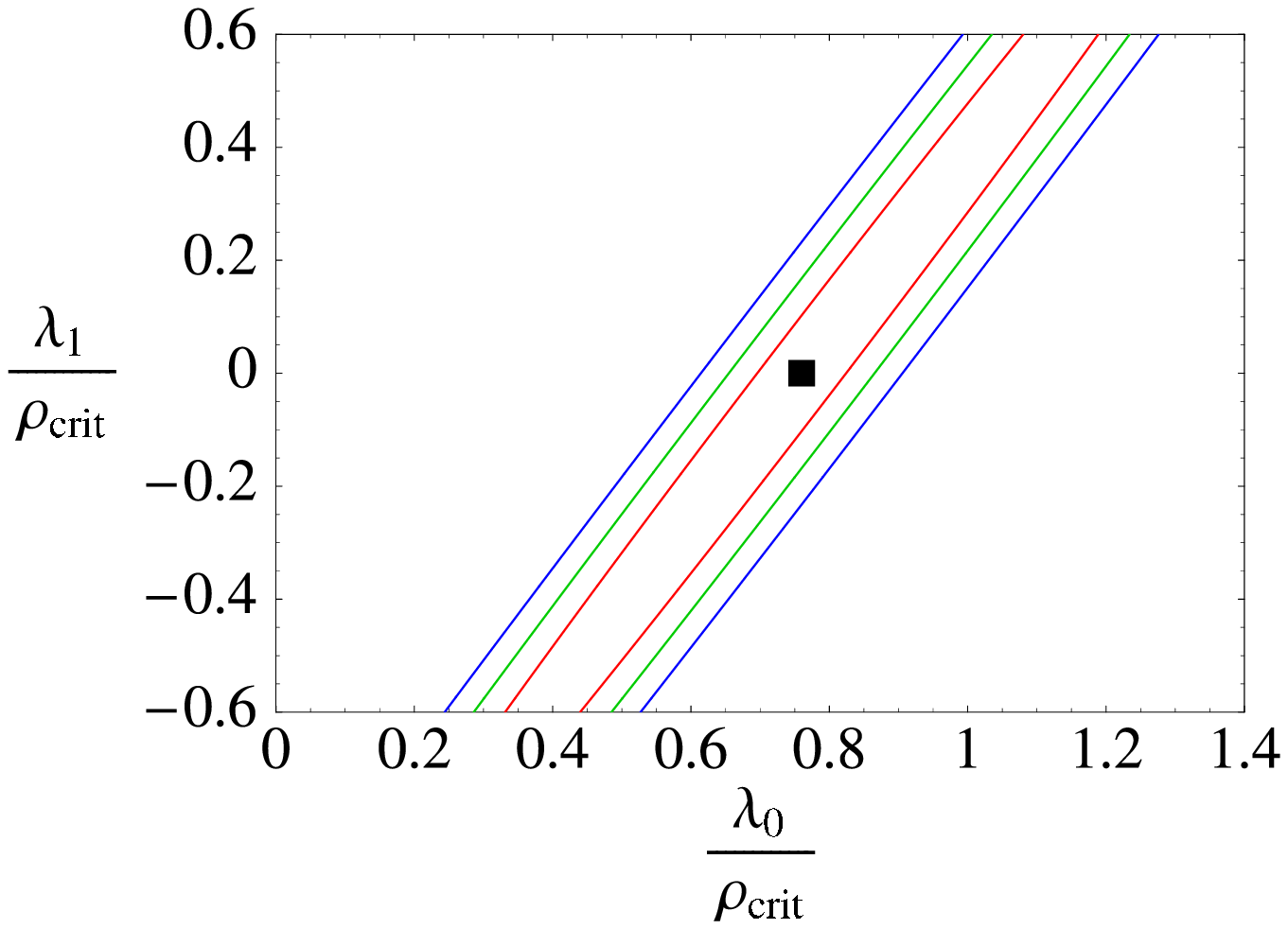} &
		 \includegraphics[width=7.5cm]{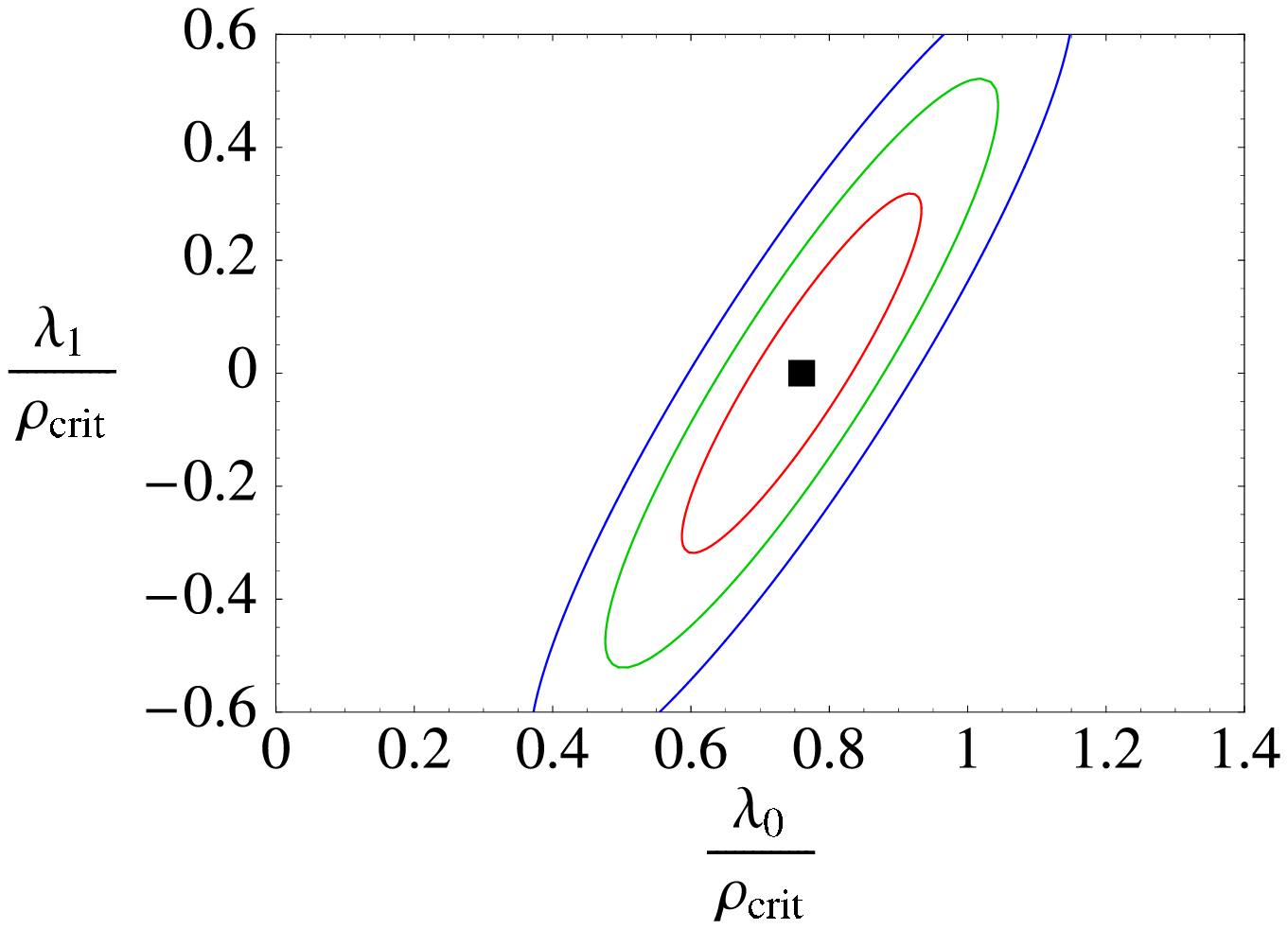} \\
\hspace{-0.55cm} \includegraphics[width=7.5cm]{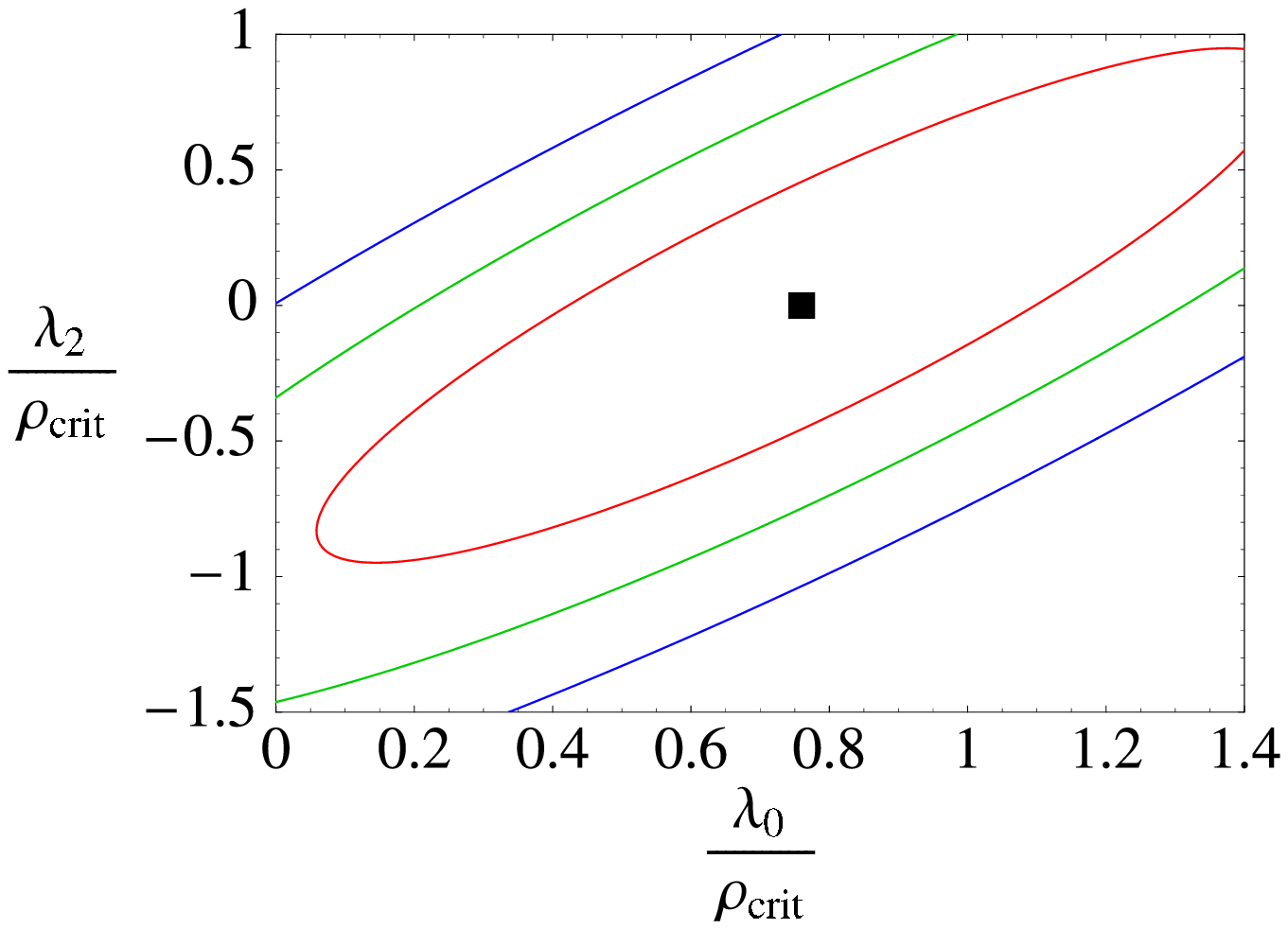} &
		 \includegraphics[width=7.5cm]{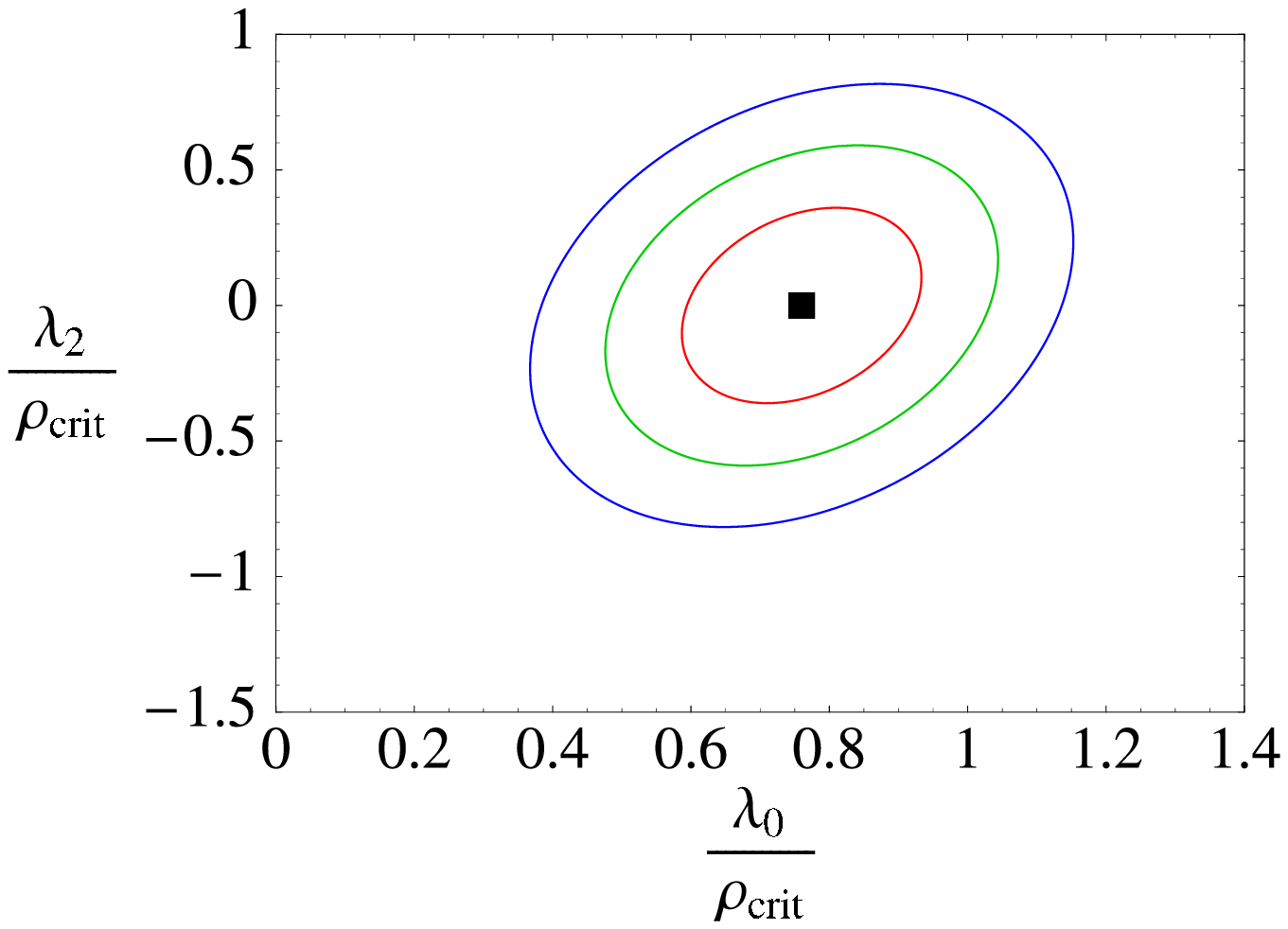}
\end{tabular}
\caption{\it 1, 2 and 3 $\sigma$ contours for $\lambda_0$, $\lambda_1$ (upper panels) and $\lambda_0$, $\lambda_2$ (lower panels) from measurements of $d_A(z)$ (left) and $H(z)$ (right) alone, for a ``ground" BAO  survey. While the fractional error in $d_A(z)$ is approximately half of the error in $H(z)$, the direct dependence of $H(z)$ on  $\lambda_i$ coefficients through Eq.~(\ref{eq:Hsq-cheby-curv})  yields stronger constraints on these parameters.}
\label{fig:Hvsda_ground}
\end{center}
\end{figure}

\begin{figure}[t]
\vspace{-0cm}
\begin{center}
\begin{tabular}{cc}
\hspace{-0.55cm} \includegraphics[width=7.5cm]{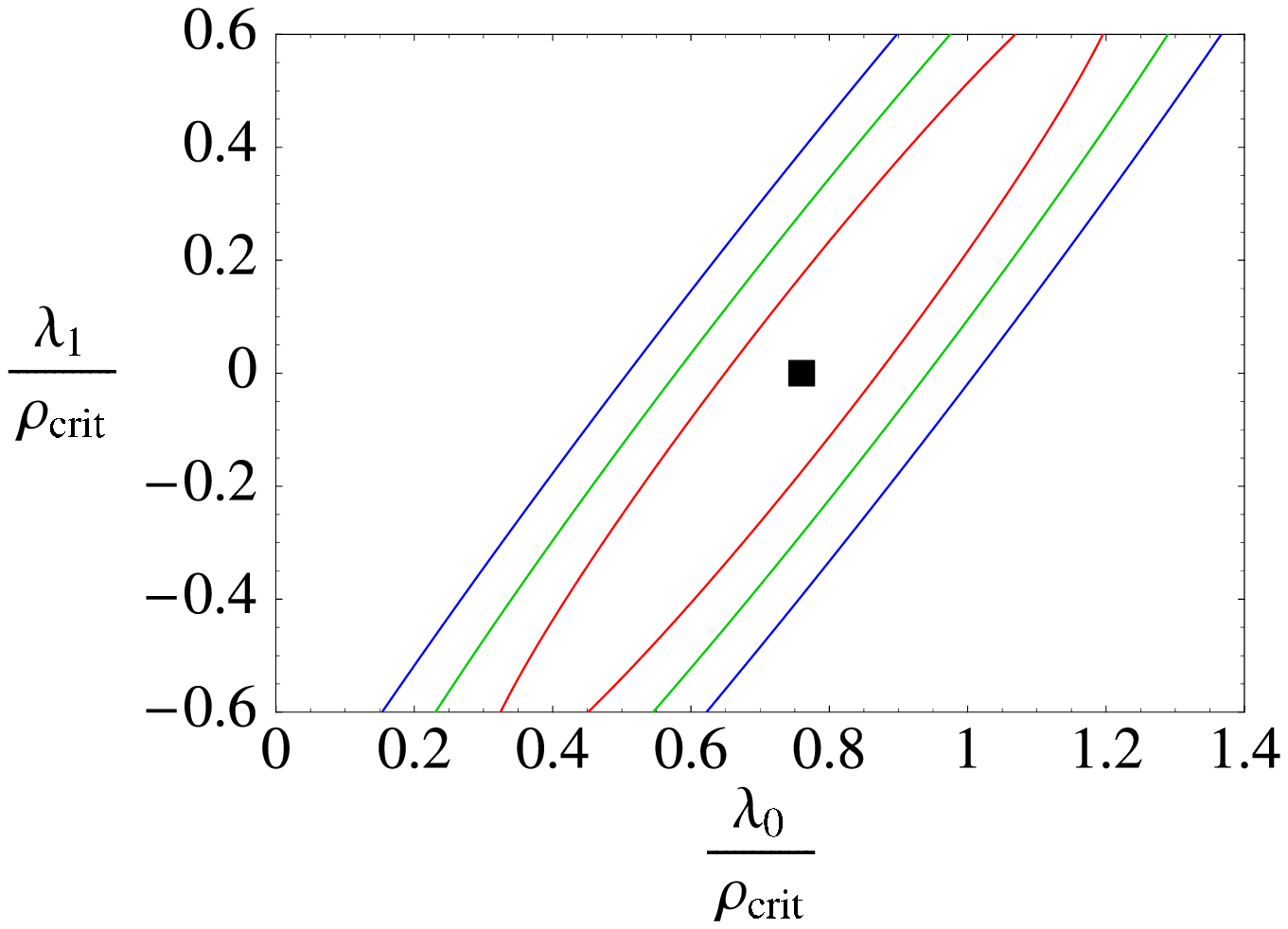} &
		 \includegraphics[width=7.5cm]{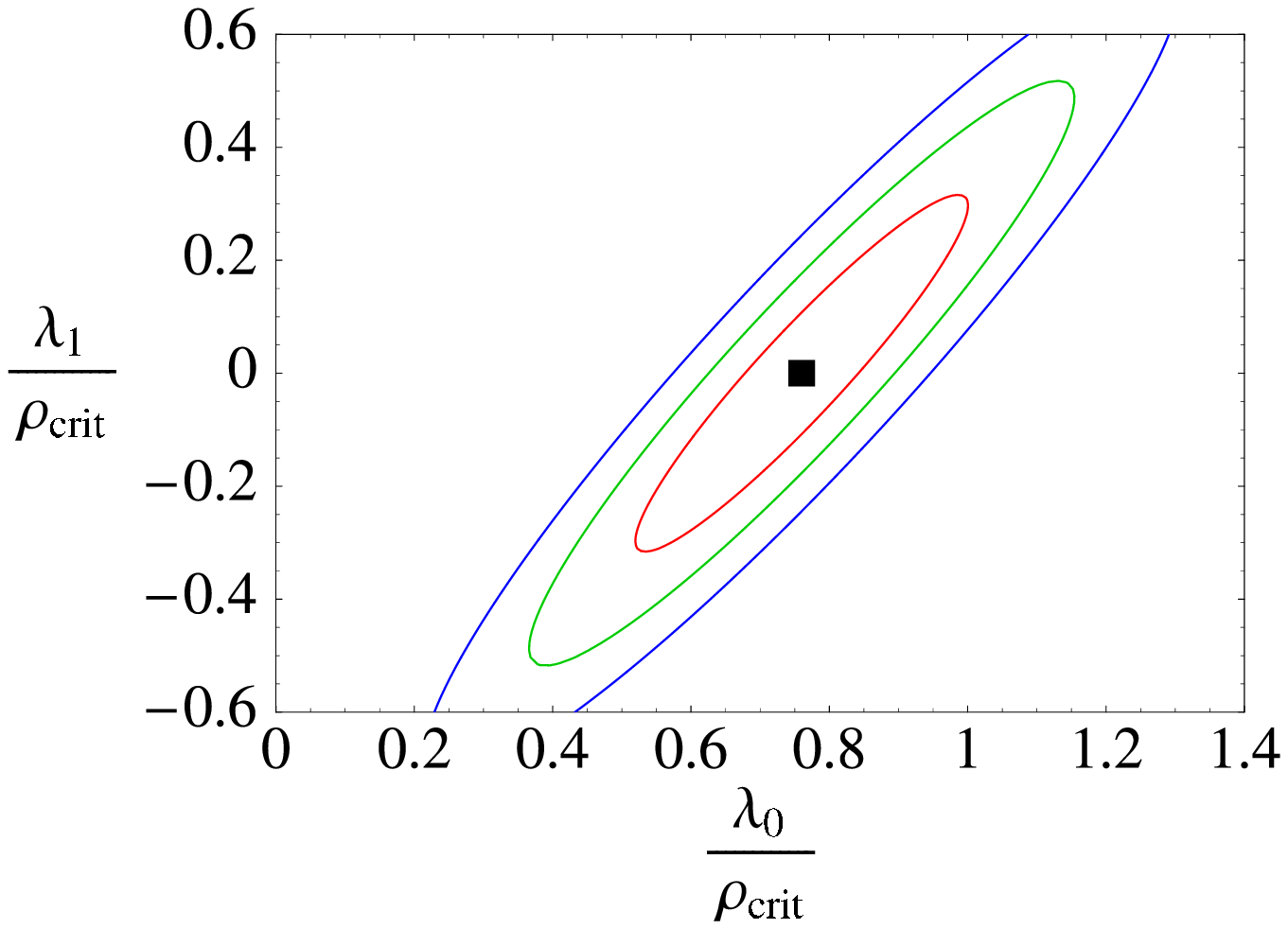} \\
\hspace{-0.55cm} \includegraphics[width=7.5cm]{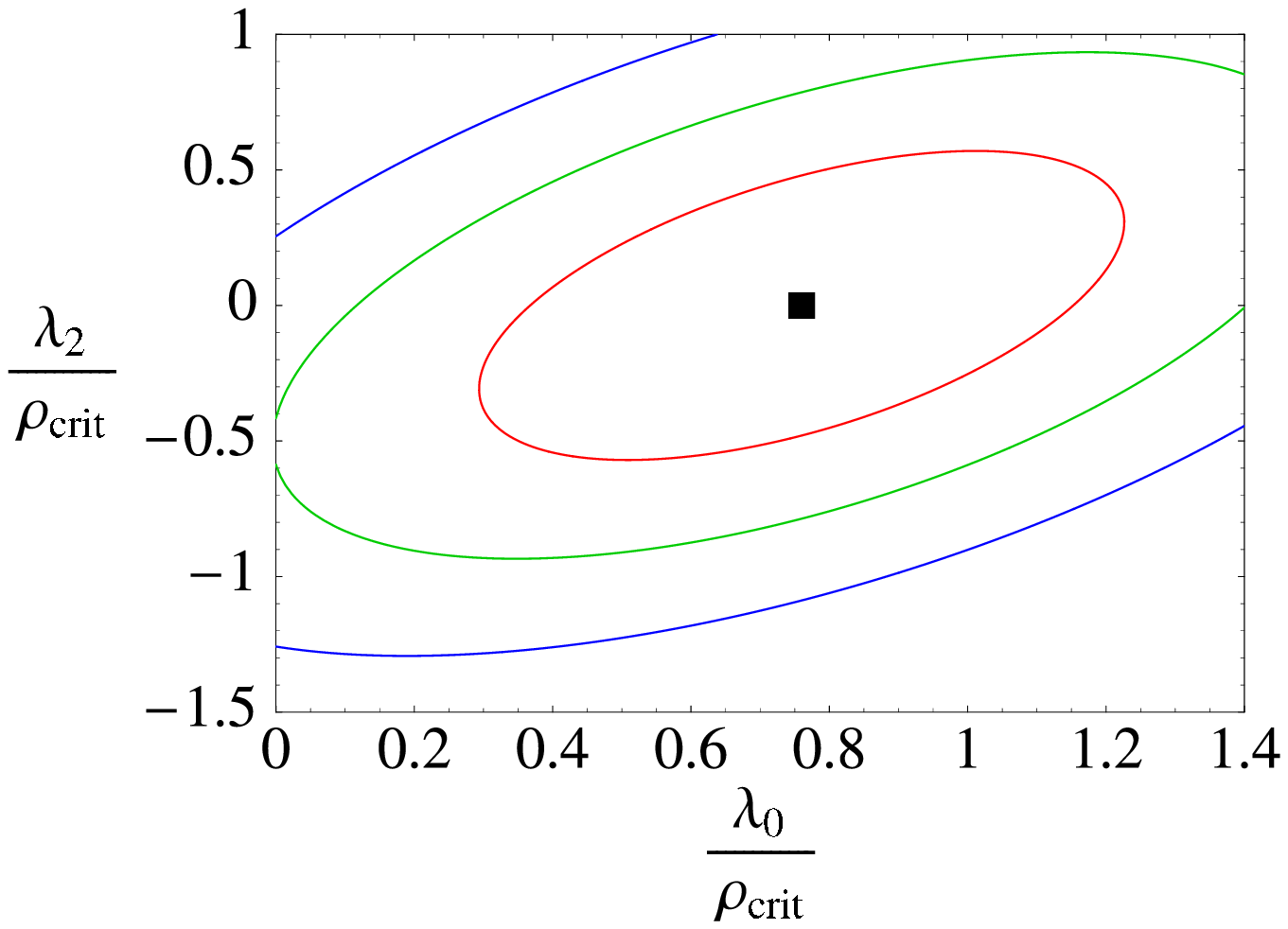} &
		 \includegraphics[width=7.5cm]{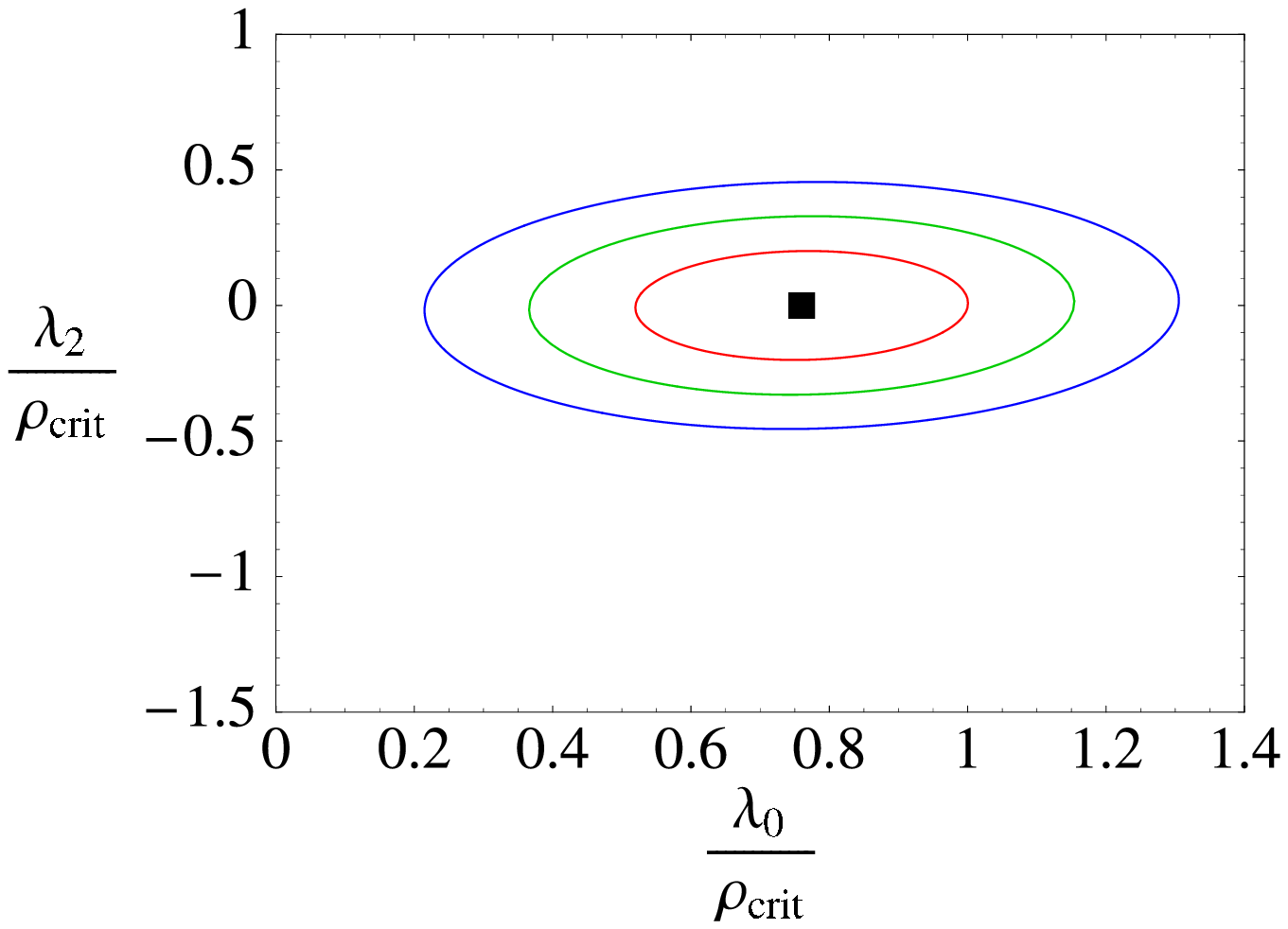}
\end{tabular}
\caption{\it 1, 2 and 3 $\sigma$ contours for $\lambda_0$, $\lambda_1$ and $\lambda_2$ from measurements of $d_A(z)$ (left) and $H(z)$ (right) alone, for a ``space" BAO  survey. While the fractional error in $d_A(z)$ is approximately half of the error in $H(z)$, the direct dependence of $H(z)$ on  $\lambda_i$ coefficients through Eq.~(\ref{eq:Hsq-cheby-curv})  yields stronger constraints on these parameters.}
\label{fig:Hvsda_space}
\end{center}
\end{figure}

It can bee seen from Eqs.~(\ref{errord}) and (\ref{errorH}) that the fractional error in $d_A(z)$ is approximately half of the error in $H(z)$. However,
there is a direct dependence of $H(z)$ on  $\lambda_i$ coefficients through Eq.~(\ref{eq:Hsq-cheby-curv}), while the relation with $d_A(z)$ involves an integral
and the bounds derived from the information on $d_A(z)$ are generally weaker than those derived from the measurement of $H(z)$. 
As an example, in Figs.~\ref{fig:Hvsda_ground} and \ref{fig:Hvsda_space},  we show the 1, 2 and 3 $\sigma$ constraints that a ``ground" or ``space'' BAO experiment could respectively place on $\lambda_0$, $\lambda_1$ and $\lambda_2$ using only the information on $d_A(z)$ (left) and on $H(z)$ (right). Notice that the constraints derived from the information on $H(z)$ are much tighter. Indeed, we found that, with information on $d_A(z)$ or $d_L(z)$ alone, there is a degeneracy between $\lambda_0$ and $\lambda_2$, as shown in the bottom left panel of Figs.~\ref{fig:Hvsda_ground} and \ref{fig:Hvsda_space} and the right panel of Fig.~4 in Ref.~\cite{Simonetal05}. This degeneracy is lifted by  data constraining $H(z)$ as can be seen in the bottom right panel of Figs.~\ref{fig:Hvsda_ground} and \ref{fig:Hvsda_space} and the right panel of Figs.~3 and 6 in Ref.~\cite{Simonetal05}. This favors spectroscopic surveys, which can measure $H(z)$, over photometric surveys with large photo-z errors which can only measure $d_A(z)$. 

Let us consider more closely  the strong degeneracy between $\lambda_0$ and $\lambda_1$ in the top panels of Figs.~\ref{fig:Hvsda_ground} and \ref{fig:Hvsda_space} and in Figs.~3, 4 and 6 of Ref.~\cite{Simonetal05}. This degeneracy is present in all the datasets we considered but it is more pronounced when no information on $H(z)$ is available and the sensitivity to the $\lambda_i$ coefficients relies in integrals like $d_A(z)$ or $d_L(z)$ as for Supernovae data. 
This degeneracy is described by a linear relation between $\lambda_0$ and $\lambda_1$ of the form:

\begin{equation}
\lambda_0 = \alpha \lambda_1 + \beta\,.
\label{linealdeg}
\end{equation}

This implies, to first order in the Chebyshev expansion of Eq. (\ref{eq:V-cheb-expan}):

\begin{equation}
V(z) = (\alpha-1)\lambda_1 + \beta + \lambda_1\frac{2z}{z_{max}}~.
\label{throat}
\end{equation}

Then, for any value of  $\lambda_1$ along this degeneracy, there is a redshift $z=z_{max}(1-\alpha)/2$ for which the value of the potential is fixed to  $V=\beta$. For all the datasets we found linear degeneracies between $\lambda_0$ and $\lambda_1$ with $\alpha \lap 1$ and $\beta \sim \Omega_\Lambda$; this means
that for all datasets the potential is better constrained at low  redshift,  in order to have  $V \sim \Omega_\Lambda$. This reflects the fact that  cosmological data are more sensitive to the dark energy properties for small $z$, since at larger redshifts the matter component dominates and the dependence on $V(z)$ is subdominant. However the exact  transition redshift (where $\Omega_m =\Omega_{DE}$) depends on the shape of the dark energy potential.

\begin{figure}[t]
\vspace{-0cm}
\begin{center}
\begin{tabular}{cc}
\hspace{-0.55cm} \includegraphics[width=7.cm]{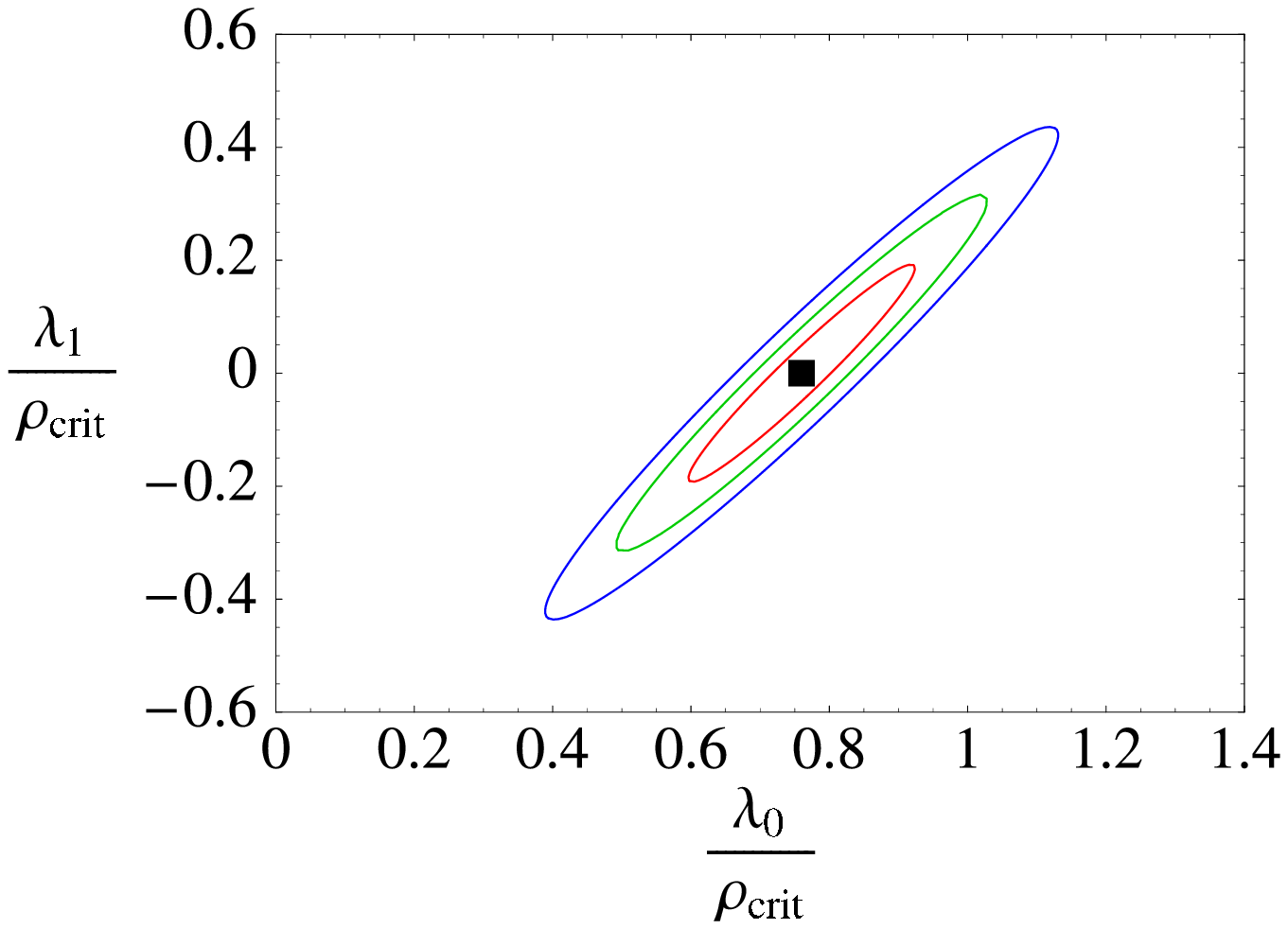} &
		 \includegraphics[width=7.cm]{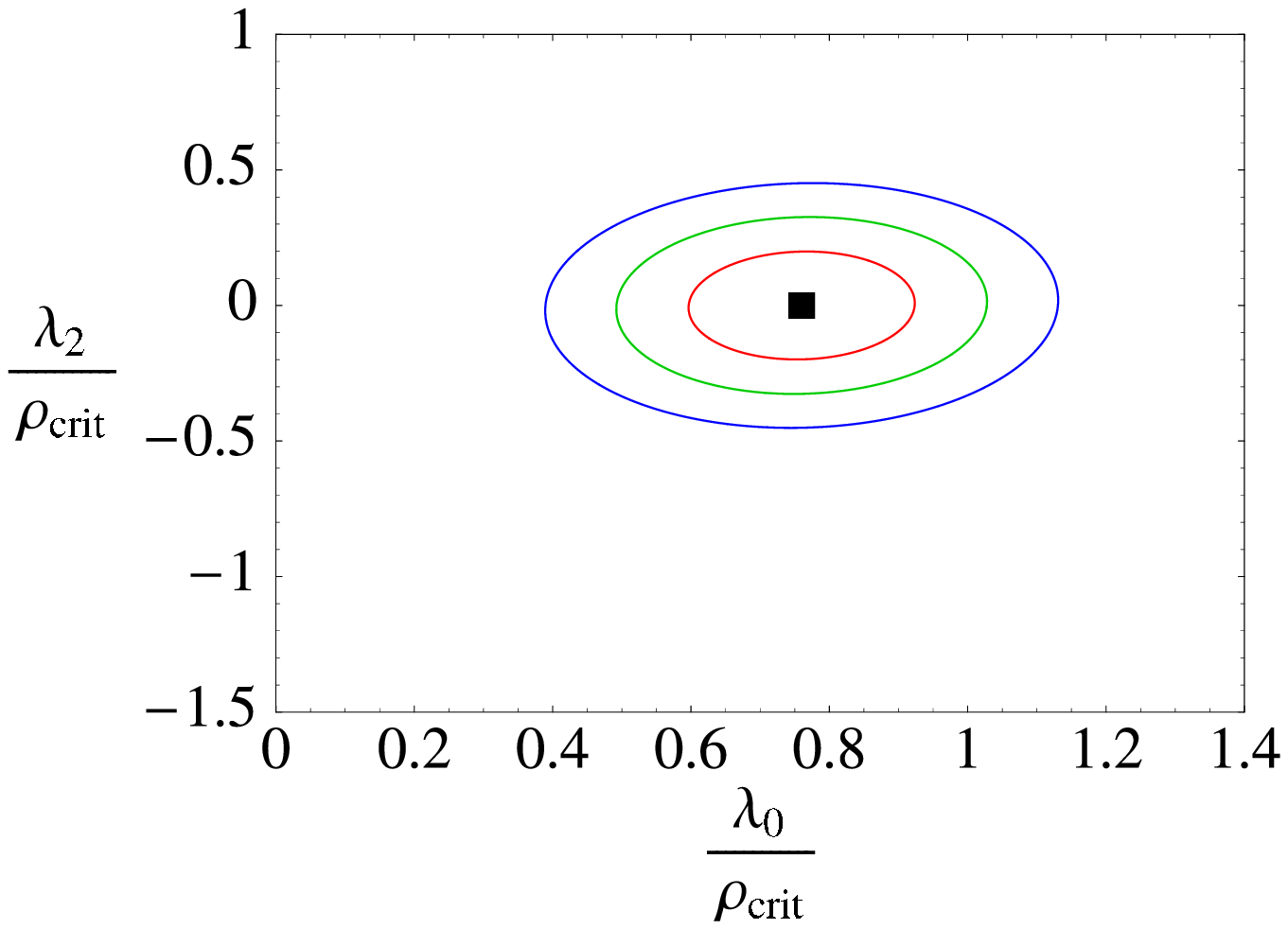} \\
\hspace{-0.55cm} \includegraphics[width=7.cm]{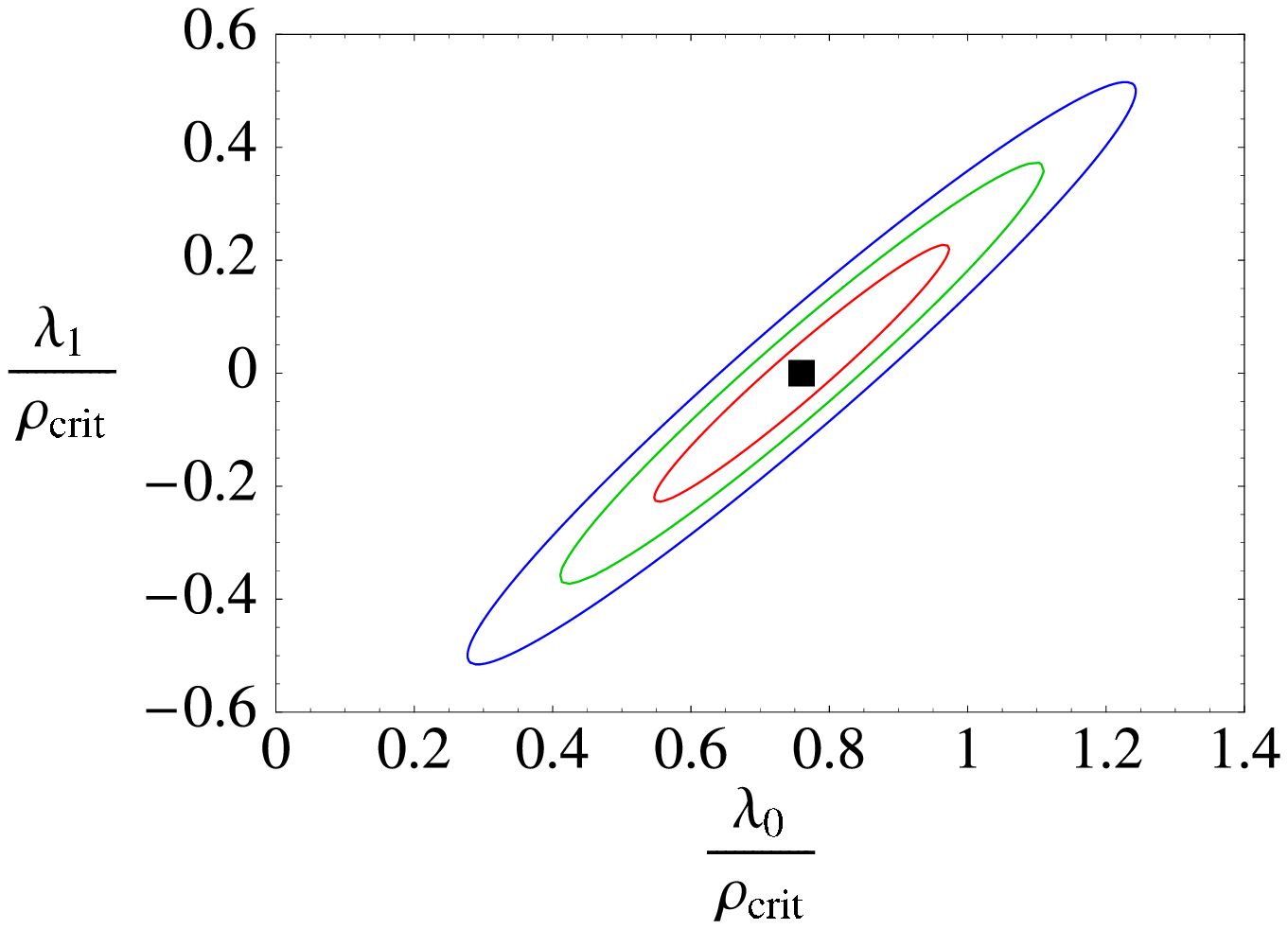} &
		 \includegraphics[width=7.cm]{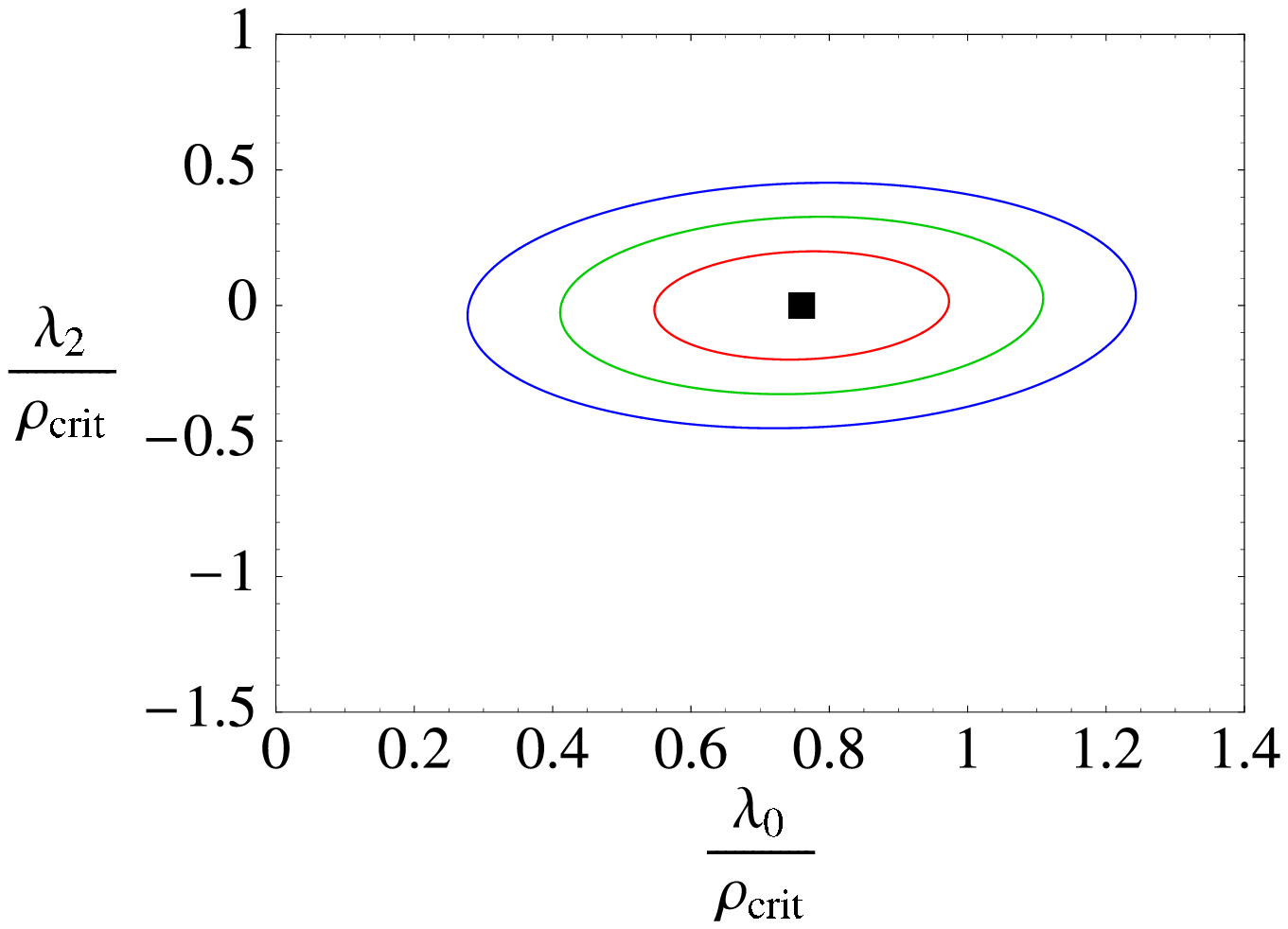} \\
\hspace{-0.55cm} \includegraphics[width=7.cm]{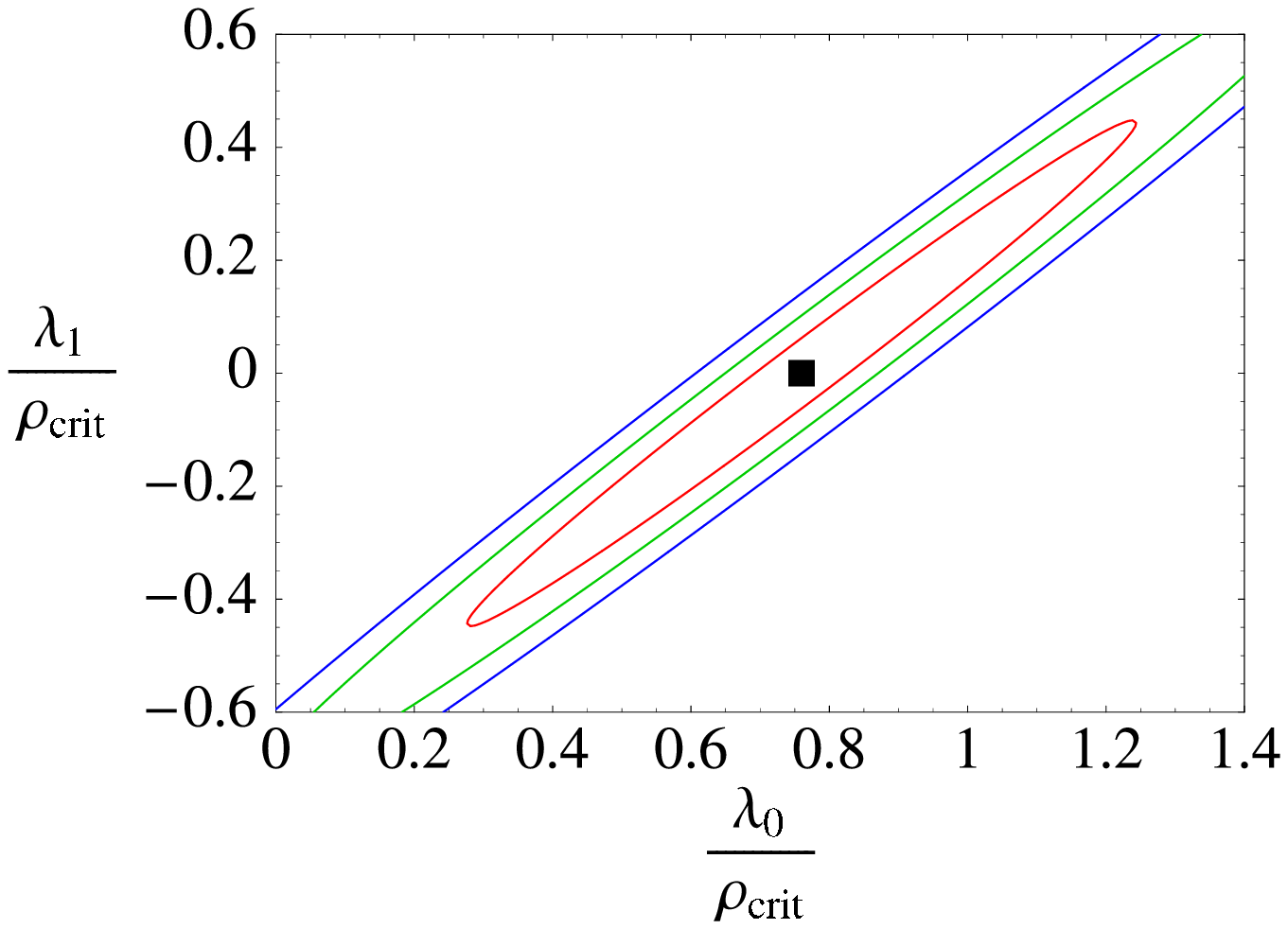} &
		 \includegraphics[width=7.cm]{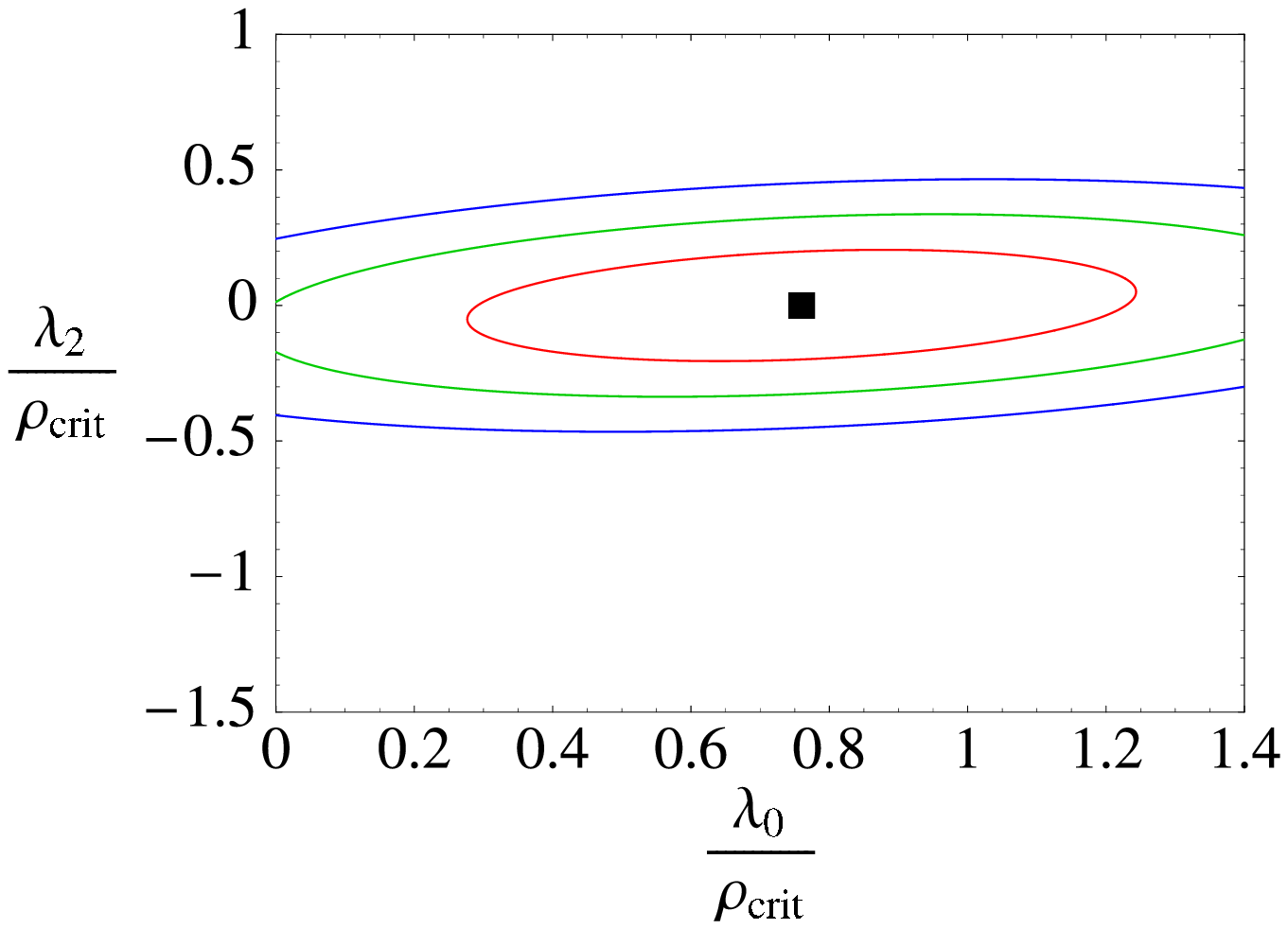} \\
\hspace{-0.55cm} \includegraphics[width=7.cm]{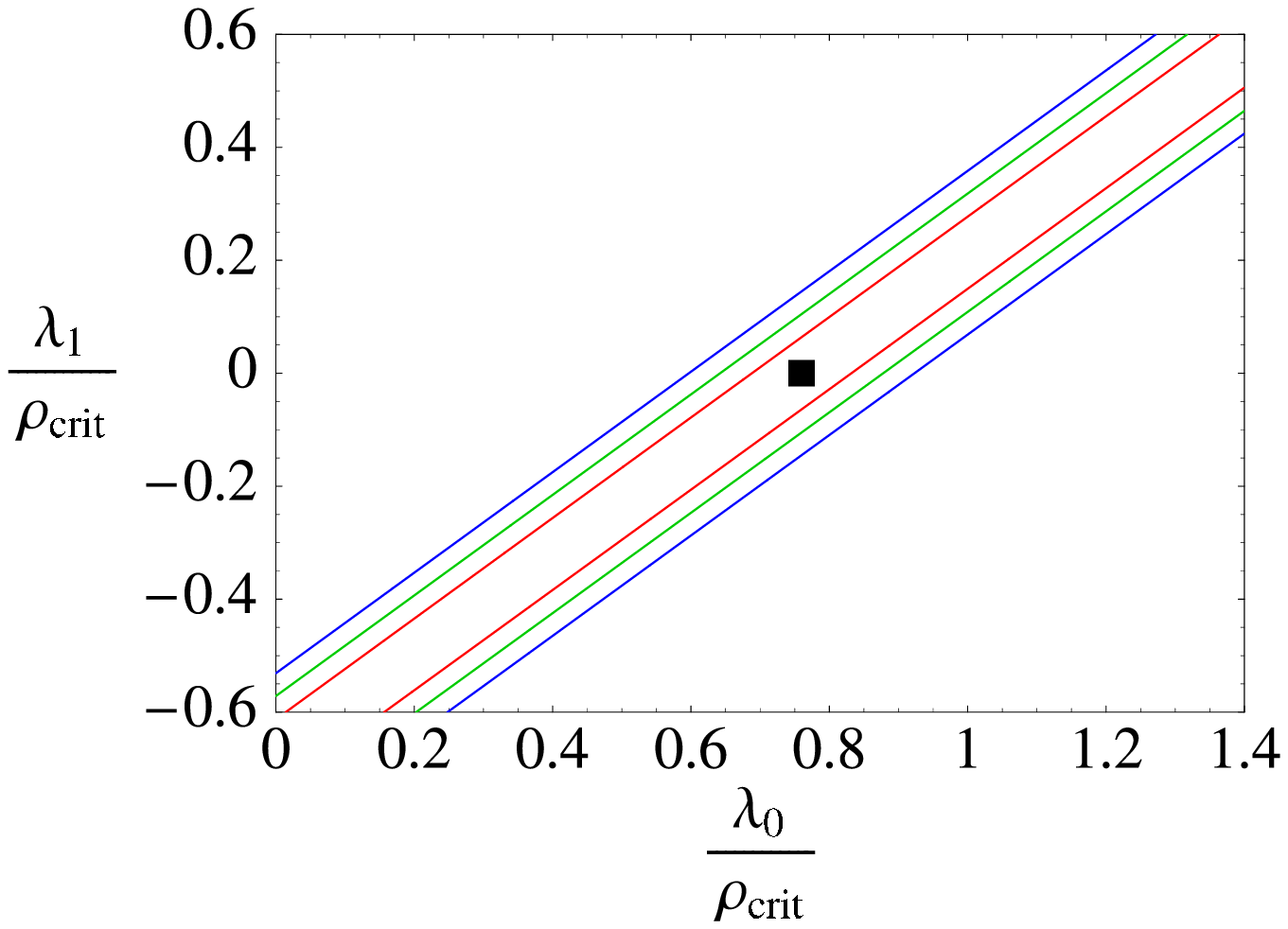} &
		 \includegraphics[width=7.cm]{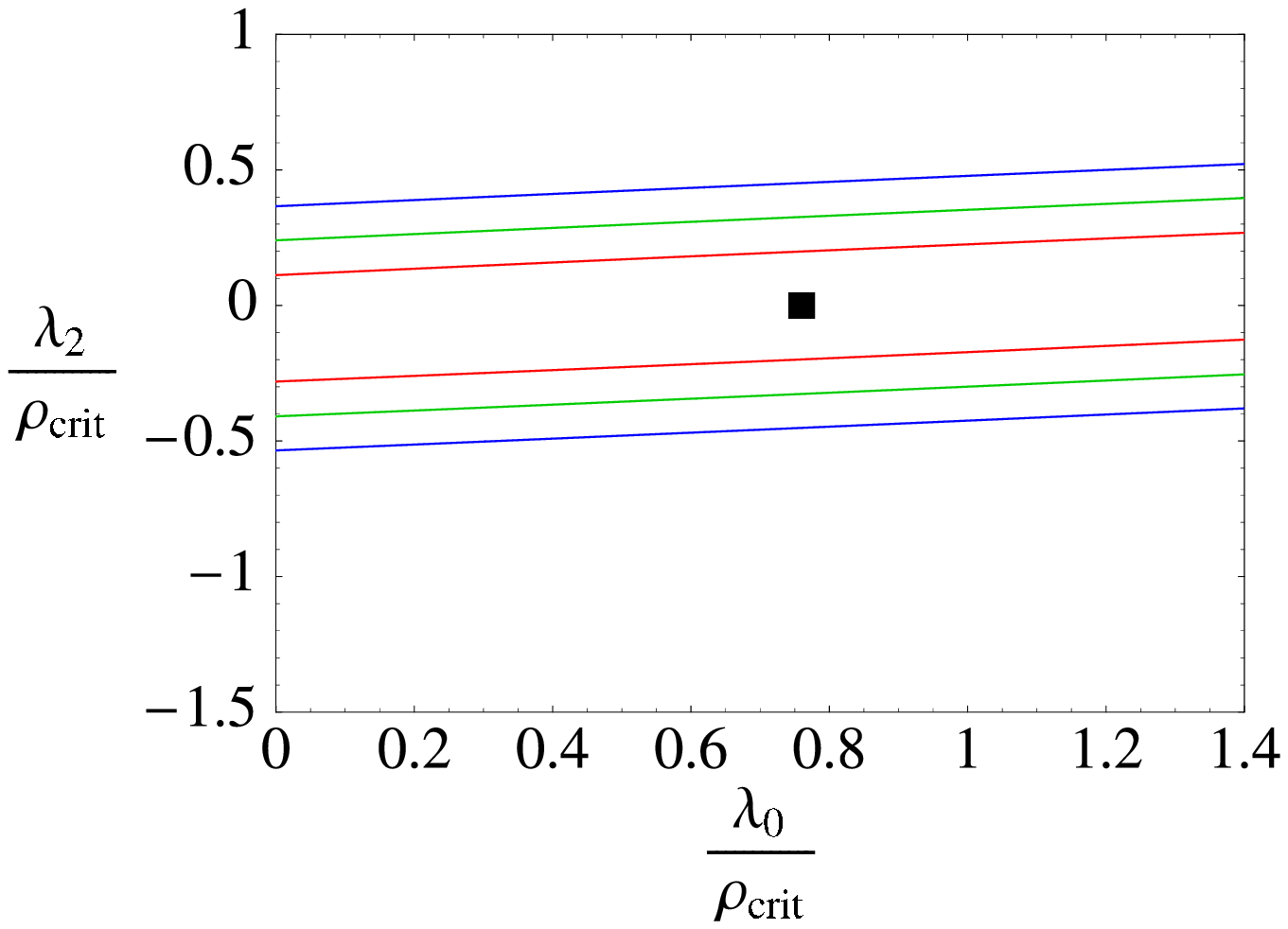} 
\end{tabular}
\caption{\it 1, 2 and 3 $\sigma$ contours for $\lambda_0$, $\lambda_1$ and $\lambda_2$  for a ``space" BAO  survey for different values of the prior on $\Omega_k$, 0, 0.03, 0.1 and 1.}
\label{fig:omegak}
\end{center}
\end{figure}

As for the effect of considering non-flat geometries we find that, with the prior of $\sigma_k=0.03$ in $\Omega_k$ we consider, the effect on the extraction of the DE properties is very small, only slightly increasing the error in the reconstructed parameters. However, loosening the prior on $\Omega_k$ can severely spoil the constraints shown here significantly worsening the degeneracies among the $\lambda_i$ coefficients. As an example we show in Fig.~\ref{fig:omegak} the effect on the constraints of the first three $\lambda_i$ by the ``space'' BAO survey for different values of the prior on $\Omega_k$, 0, 0.03, 0.1 and 1.

%\newpage

Notice that, in spite of the lack of data for $z<1$, the ``space"-type survey is placing strong constraints also in that redshift region. This is a consequence of  {\it a) } the priors imposed at  $z=0$, {\it b)} the very accurate data for $z>1$ and {\it c)} the information on $w(z)$ enclosed in the $d_A(z)$ constraints, making it possible to  interpolate  the potential given by our smooth parameterization to the low redshift 
regime. Therefore, the combination of the two surveys does not provide a significant improvement of the constraints on $V(z)$ over the ``space" survey alone, however these constraints are much more robust since both experiments now cover the whole redshift range shown  in Fig.~\ref{fig:potential}. 

\begin{figure}[t]
\vspace{-0.5cm}
\begin{center}
\begin{tabular}{cc}
\hspace{-0.55cm} \includegraphics[width=7.5cm]{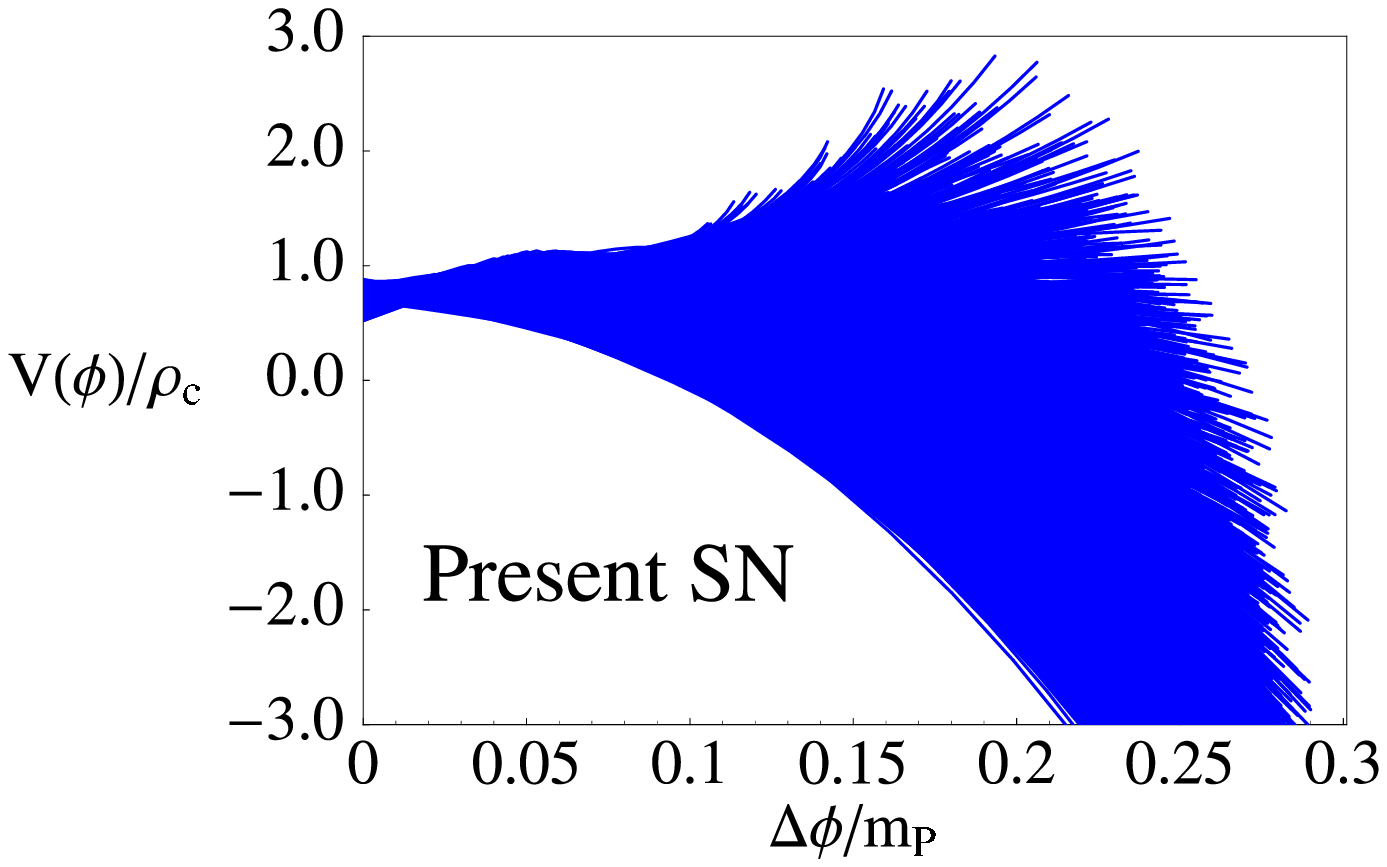} &
		 \includegraphics[width=7.5cm]{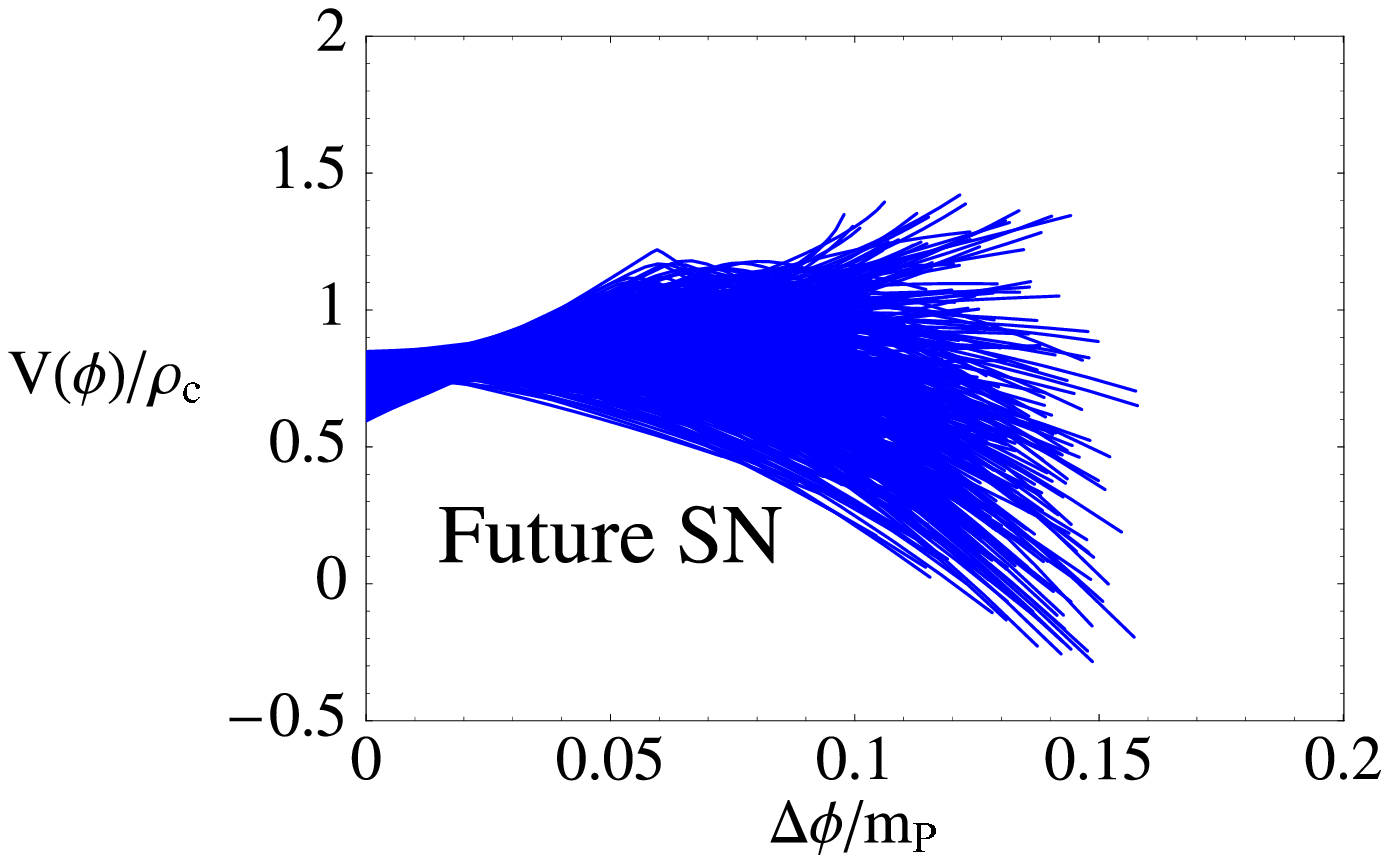} \\
\hspace{-0.55cm} \includegraphics[width=7.5cm]{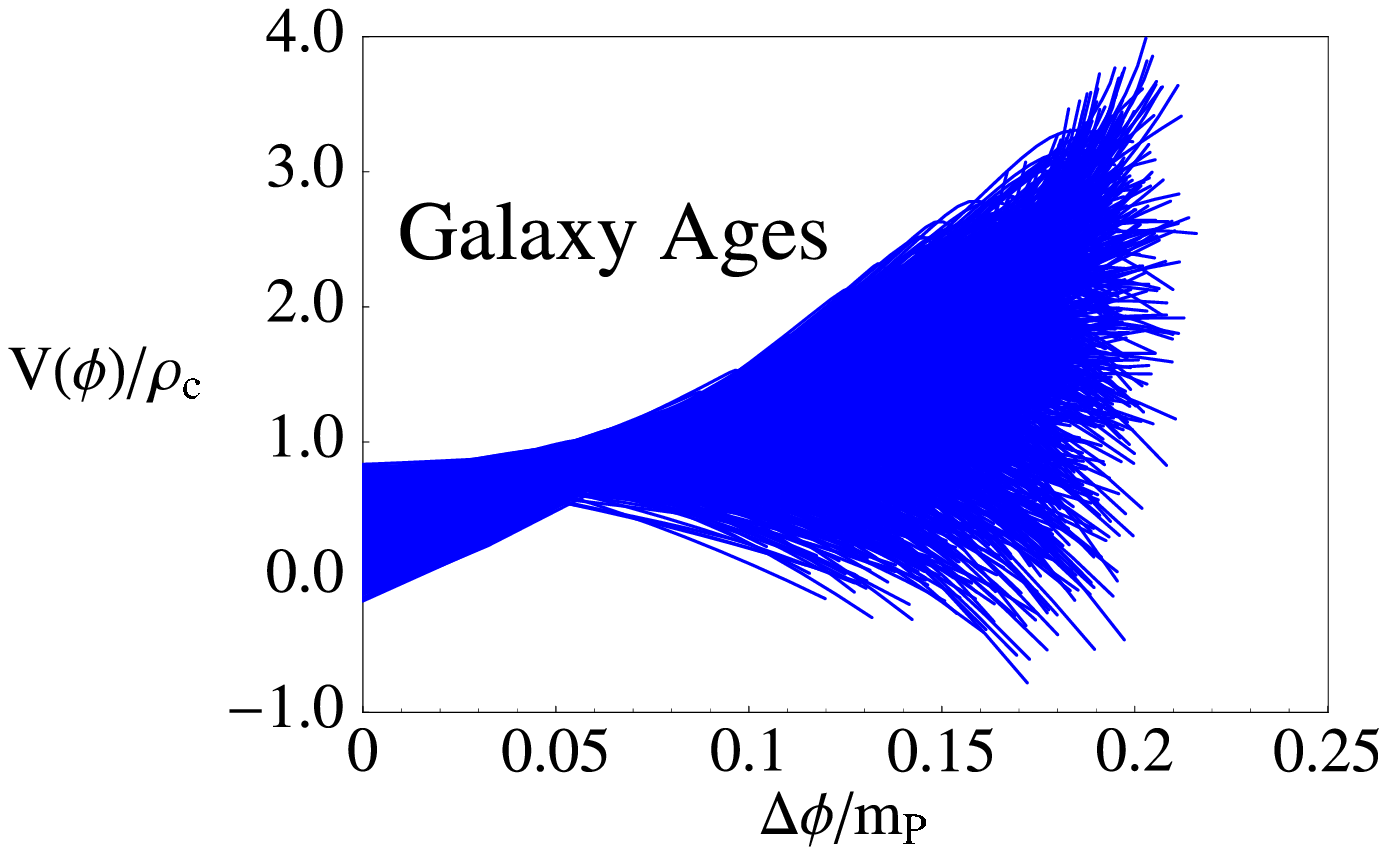} &
		 \includegraphics[width=7.5cm]{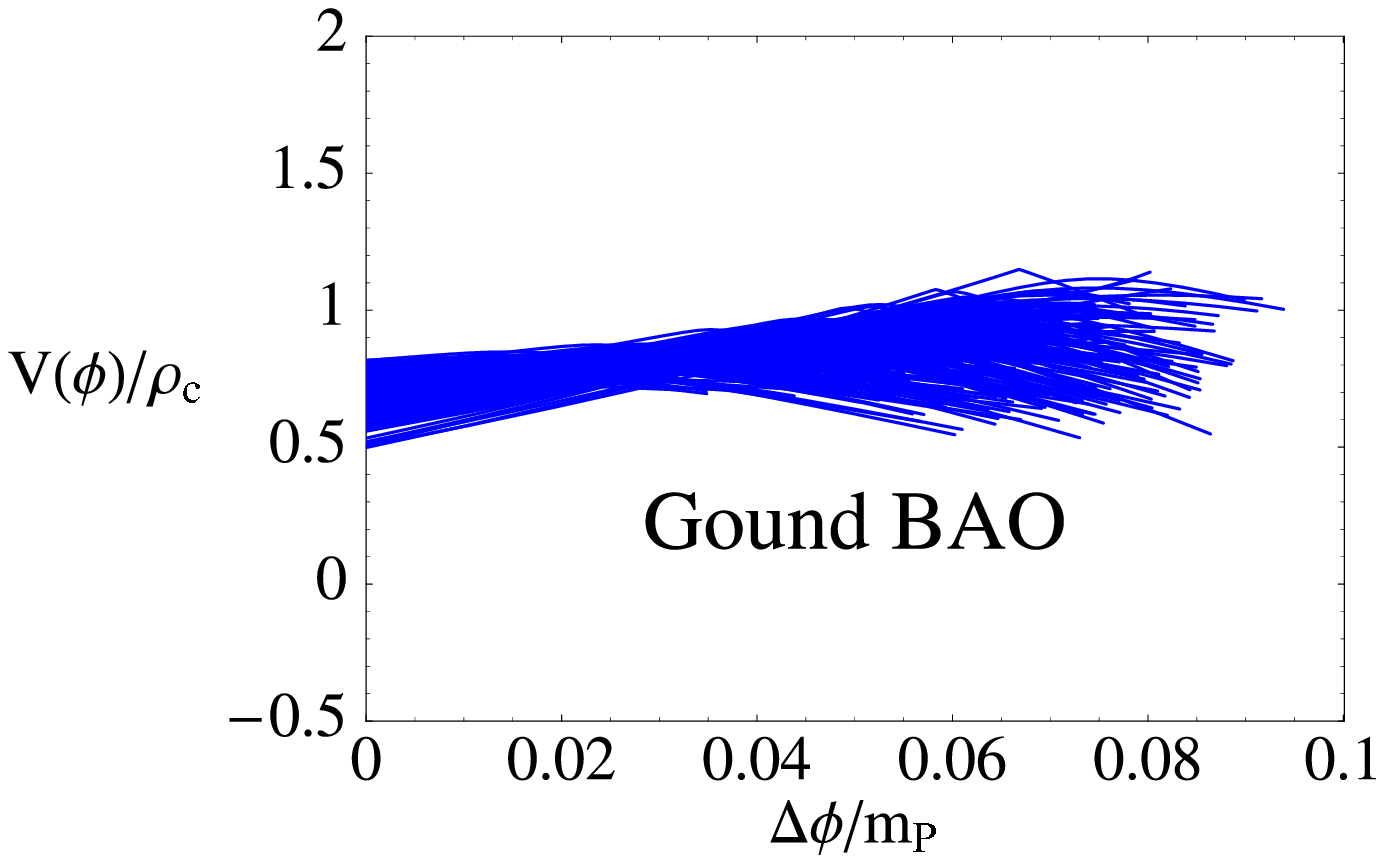} \\
\hspace{-0.55cm} \includegraphics[width=7.5cm]{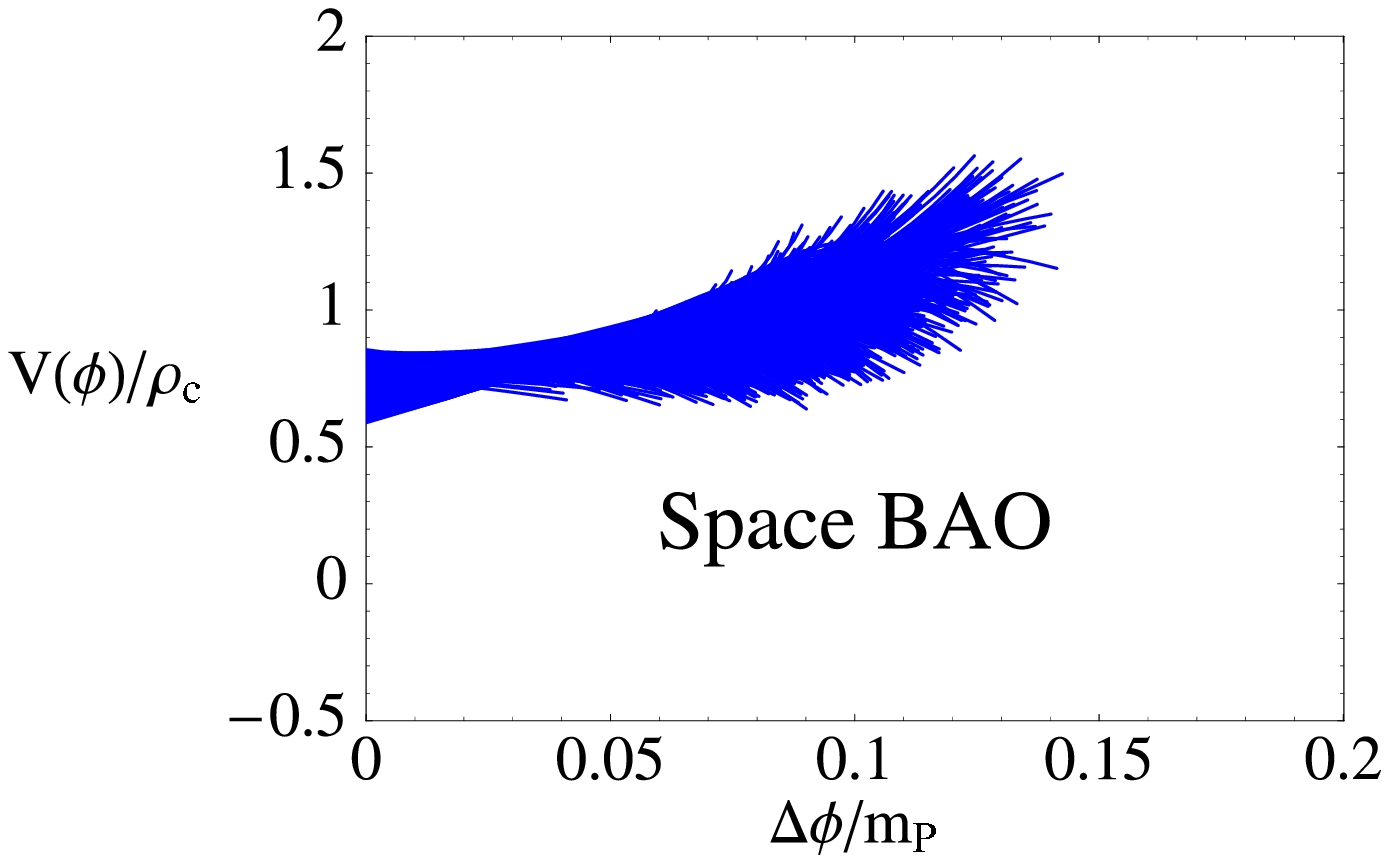} &
		 \includegraphics[width=7.5cm]{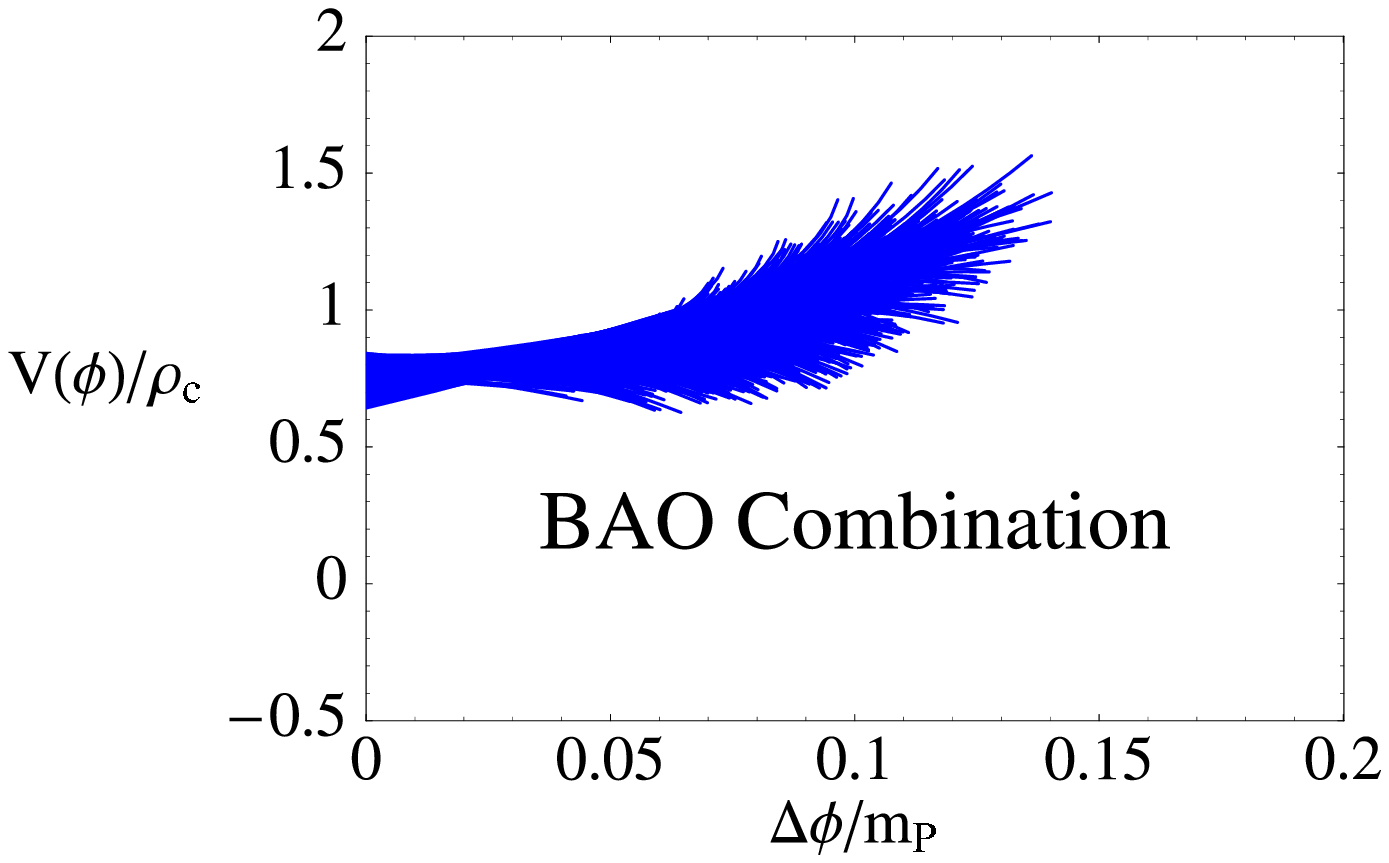}
\end{tabular}
\caption{\it  Instead of reporting the 1 and 2 $\sigma$ contours, we plot the reconstructed $V(\phi)$ for the $68\%$ best models for the different datasets (see text for more details).}
\label{fig:potentialphi}
\end{center}
\end{figure}

\subsection{Reconstructed $V(\phi)$}
Eq.~(\ref{eq:qzsol}) also enables us to reconstruct $\Delta \phi(z)$ from $V(z)$ thus constraining $V(\phi)$. Fig.~\ref{fig:potentialphi} shows the results of such a reconstruction for the $68\%$ best models for the different datasets. Note that, upon integration of Eq.~(\ref{eq:qzsol}) up to $z_{max}$, a range of $\Delta \phi(z)$ can be obtained up to a maximum value when $z=z_{max}$. This maximum value will strongly depend on the actual model that is integrated and on how strongly the field evolves in that model. Thus, not all values for $\Delta \phi$ are allowed and showing the 1 and 2 $\sigma$ contours would not be fully correct. Indeed, for the $\Lambda$CDM Eq.~(\ref{eq:qzsol}) will always yield $\Delta \phi(z)=0$ regardless of $z_{max}$. If the constraints placed by a given data set on the model are tightly centered around the $\Lambda$CDM very small values of $\Delta \phi(z)$ will be recovered from such models. 

\section{Reconstruction of the Equation of State}

It is widespread to parameterize dark energy not by the scalar field potential but by its equation of state. 
As  long as the equation of state $w$ is $> -1$, parameterizing the dynamics of dark energy through the evolution of its effective equation of state is equivalent to considering the redshift evolution of the dark energy potential.  However if we want to allow $w < -1$ then the scalar field description as presented above fails. 

Thus, considering the evolution of an effective equation of state is more general than considering the potential of a scalar field. In this case it is easier to relate $w(z)$ with the observables. Expanding the redshift dependence of the equation of state in Chebyshev polynomials analogously to the expansion of $V(z)$:
\begin{equation}
w(z) \simeq \sum_{i=0}^{N}\omega_iT(x(z))
\label{eq:w_cheby}
\end{equation}
and substituting in the first Friedmann equation we would have:
\begin{eqnarray}
  H^2(\omega_i,z)\simeq H_0^2\left[\Omega_{m,0}(1+z)^3 + \Omega_{k,0}(1+z)^2\right. \nonumber \\
 \left.+ (1-\Omega_{m,0}-\Omega_{k,0})(1+z)^3\exp\left(\frac{3}{2}z_{max}\sum_{n=0}^N
  \omega_n G_n(z)\right) \right]~,
 \label{eq:Hzwcheby}
\end{eqnarray}
where now
\begin{equation}
  G_i(z)= \int_{-1}^{2z/z_{max}-1} T_i(x)(a+bx)^{-1}dx~.
 \label{eq:G(z)}
\end{equation}

Note that in this parameterization the present-day value of $w$ is
given by
\begin{equation}
w_0=\sum_{i=0}^N(-1)^i\omega_i
\label{eq:wtoday}
\end{equation}

Eqs. (\ref{eq:w_cheby}--\ref{eq:G(z)}) are a generalization of sec 3 of \cite{Simonetal05} to non flat geometries.

\begin{figure}[t]
\vspace{-0.5cm}
\begin{center}
\begin{tabular}{cc}
\hspace{-0.55cm} \includegraphics[width=7.5cm]{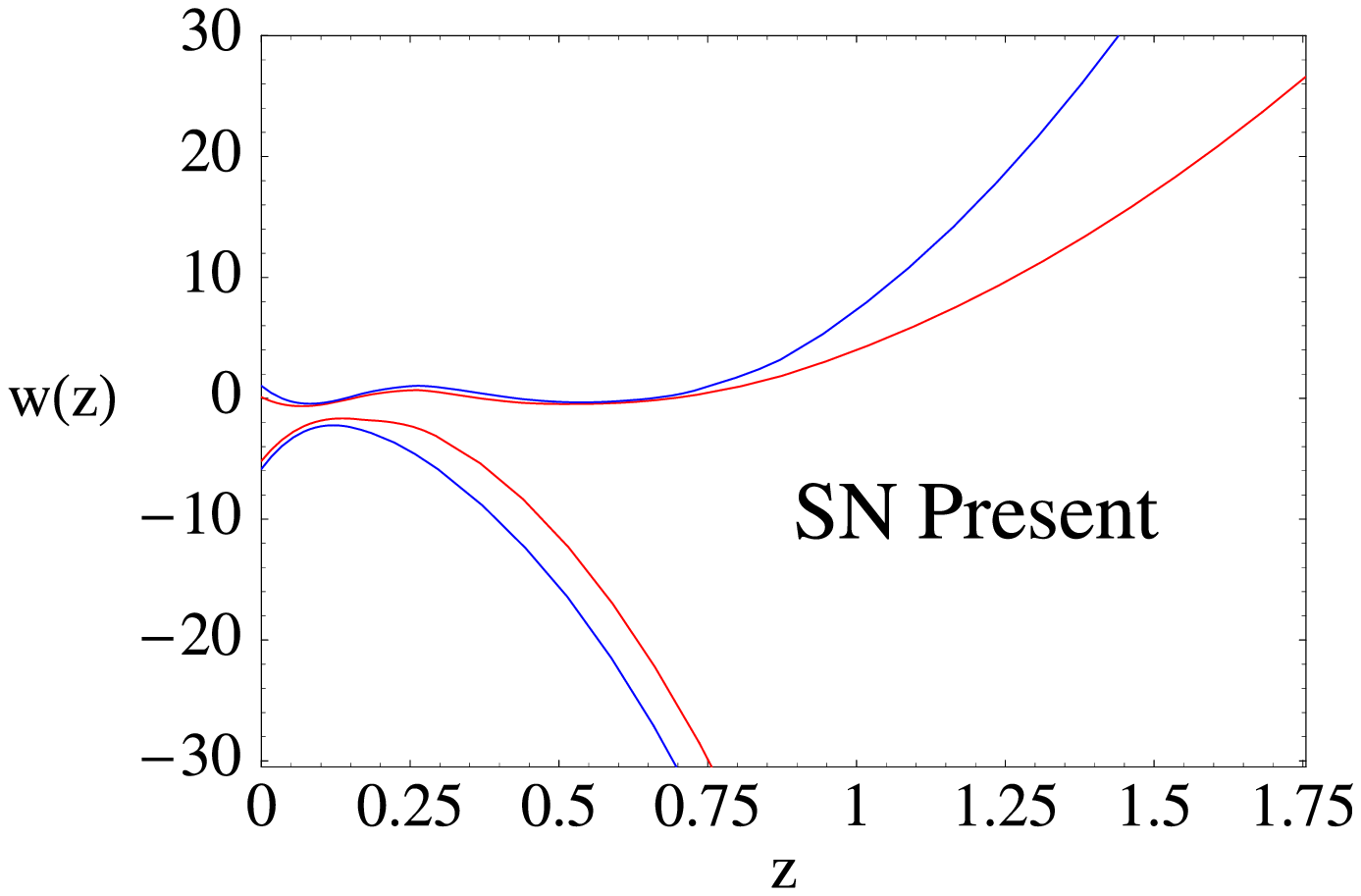} &
		 \includegraphics[width=7.5cm]{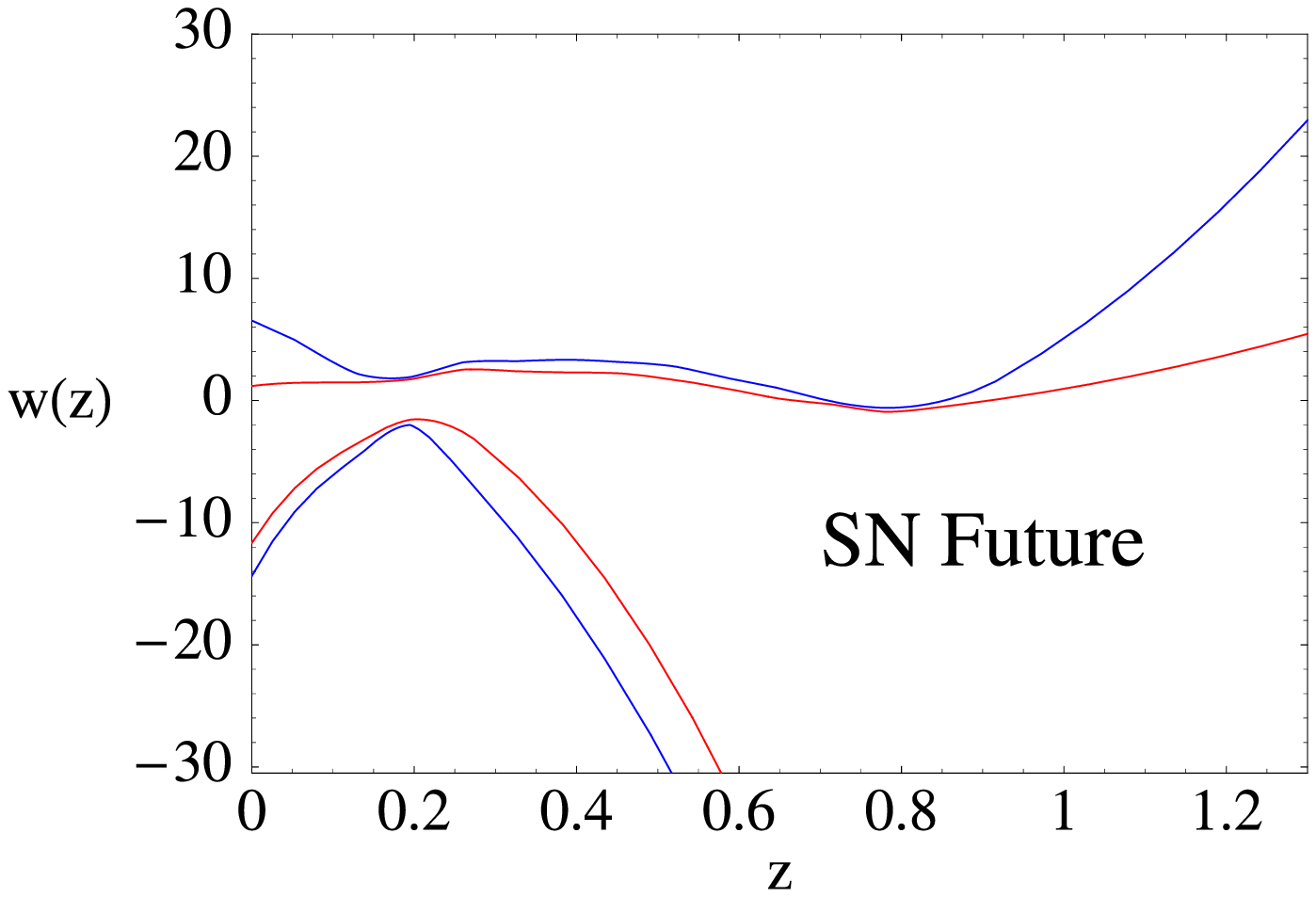} \\
\hspace{-0.55cm} \includegraphics[width=7.5cm]{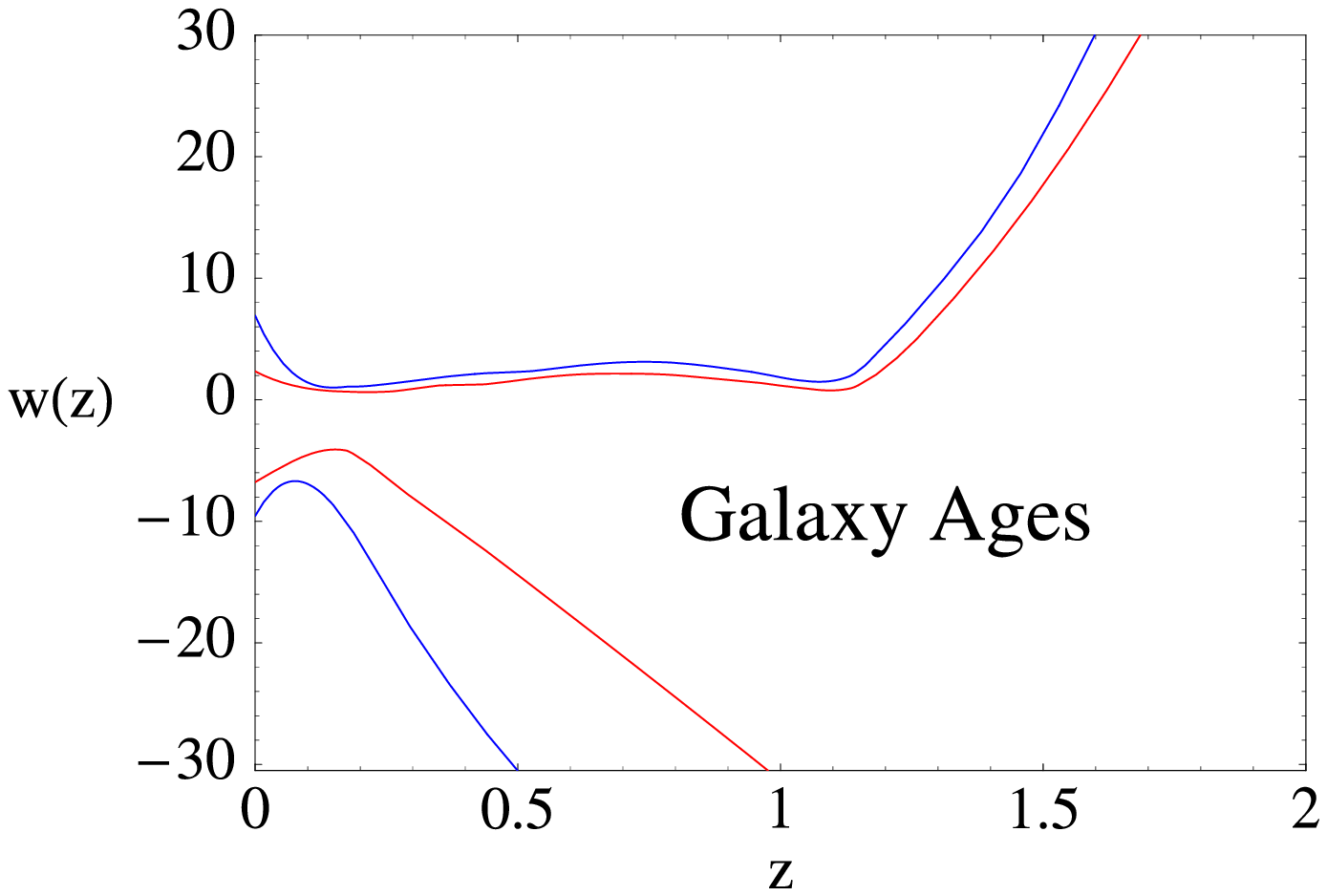} &
		 \includegraphics[width=7.5cm]{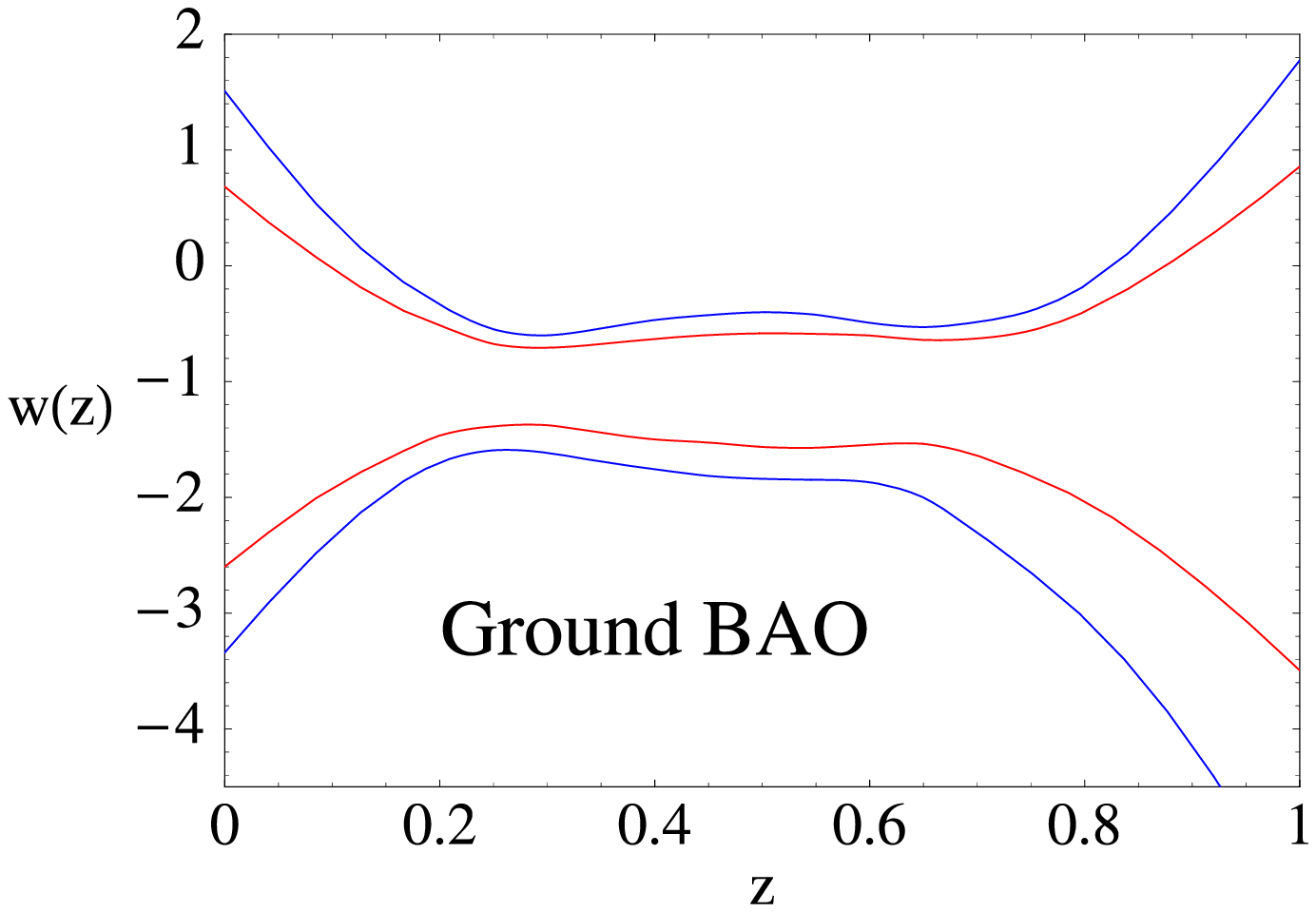} \\
\hspace{-0.55cm} \includegraphics[width=7.5cm]{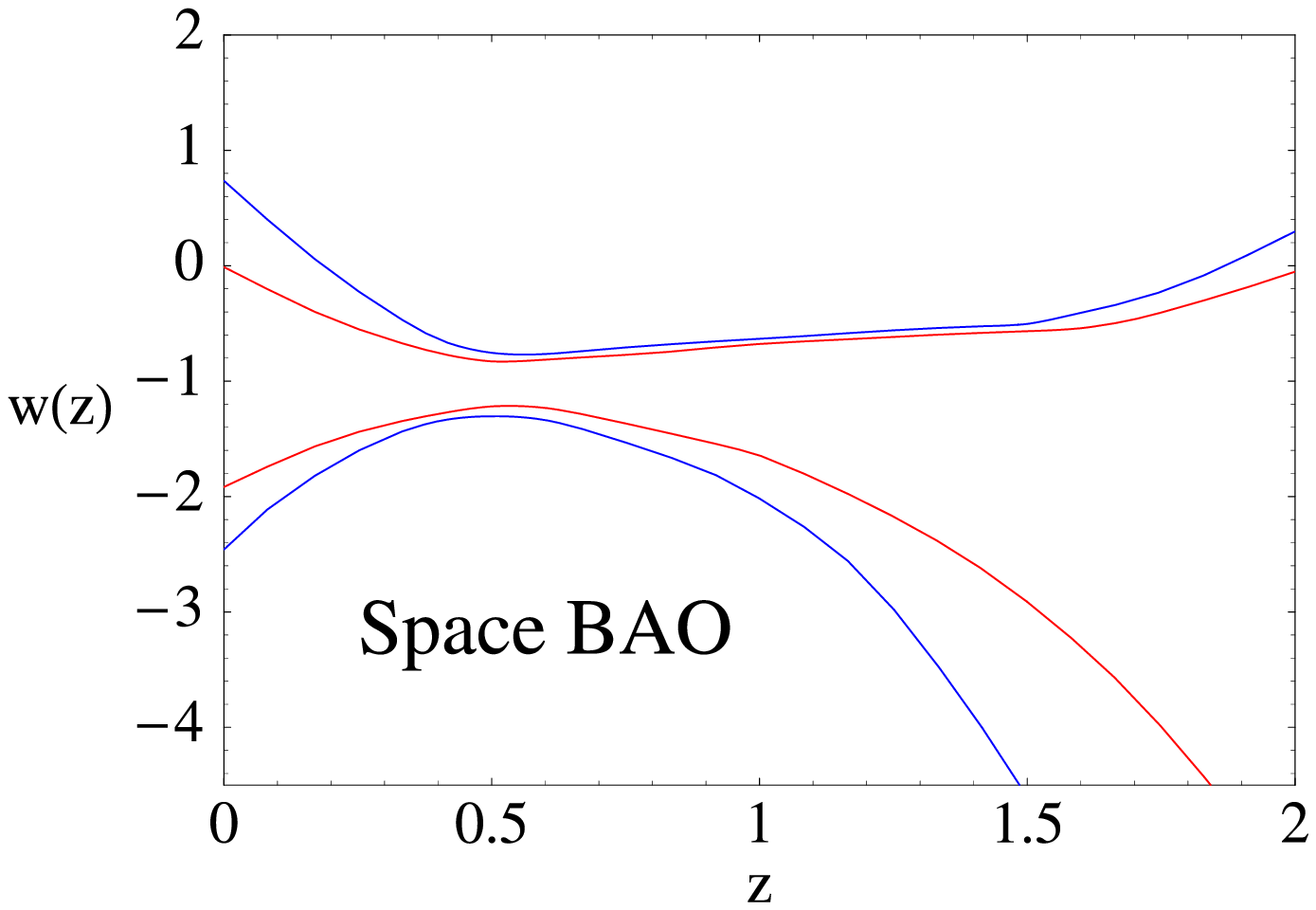} &
		 \includegraphics[width=7.5cm]{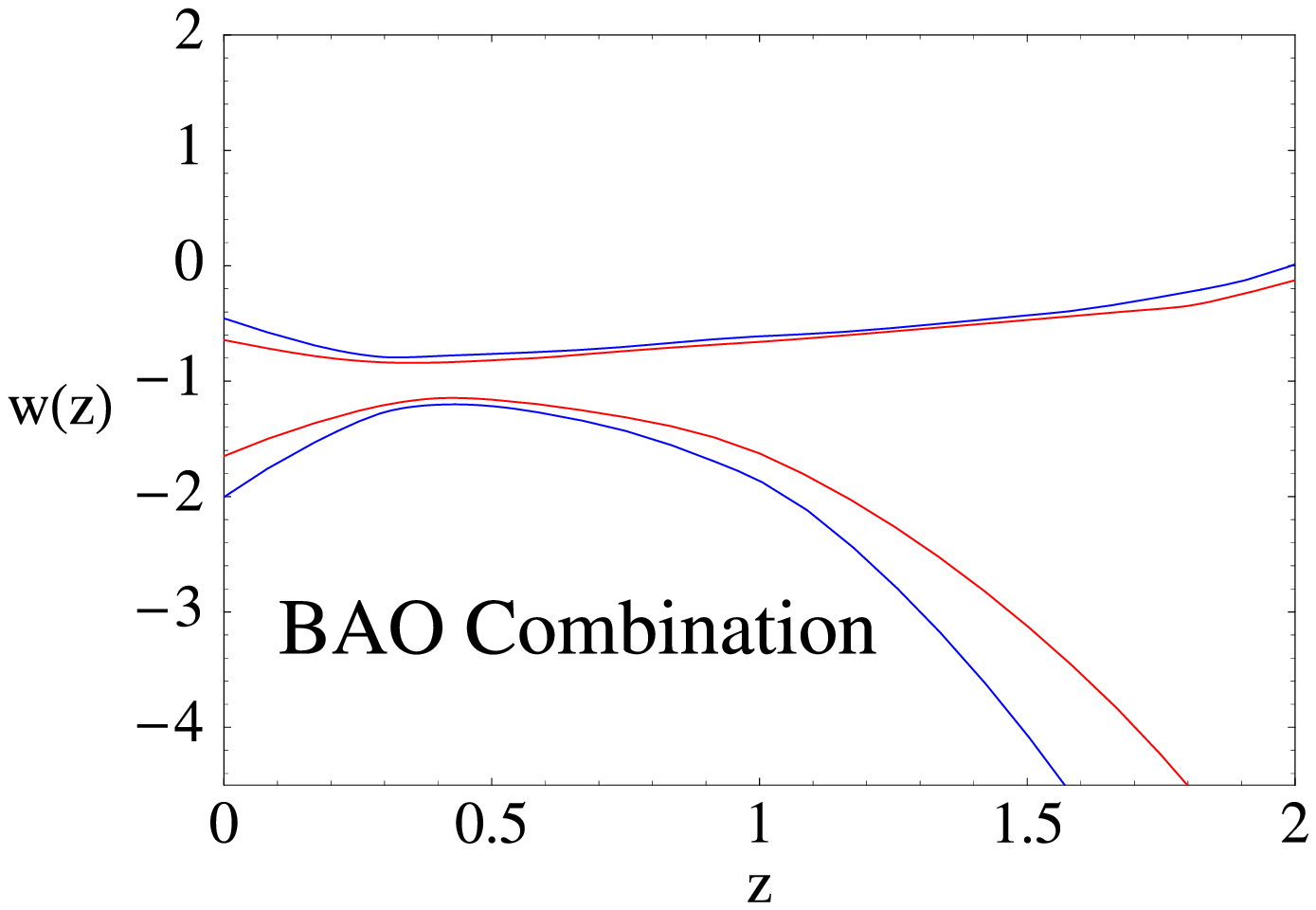}
\end{tabular}
\caption{\it 1 and 2 $\sigma$ constraints on the DE equation of state from present SN data (top left), SN data from a future-space-based experiment (top right), galaxy ages (middle left), ``ground" BAO survey (middle right), ``space" BAO survey  (bottom left) and the combination of the two BAO surveys (bottom right).}
\label{fig:eos}
\end{center}
\end{figure}

We study the constraints that the  datasets above can put on the DE equation of state through (\ref{eq:Hzwcheby}) expanding the dark energy equation state up to second order in Chebyshev polynomials. We perform forecasts using MCMC's. As in Sec.~\ref{sec:results}, $H_0$, $\Omega_m$ and $\Omega_k$ are parameters  which we marginalize over.
The same priors quoted before are also assumed here. We show the results in Fig.~\ref{fig:eos}. As for the reconstruction of $V(z)$, dark energy properties are best constrained at $z \lap 0.3$ and the  $H(z)$ determination is crucial in constraining the dark energy  evolution especially for non-trivial deviations from a constant equation of state parameter.

\begin{figure}[t]
\vspace{0.5cm}
\begin{center}
\hspace{-0.55cm} \includegraphics[width=7.5cm]{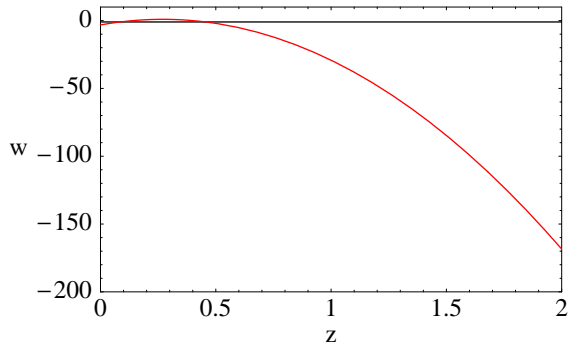} 
\caption{\it  Equation of state parameter for a model lying along the degeneracy given by dataset that can only constrain $d_A$ or $d_L$. For comparison also the $\Lambda$CDM case is shown ($w=-1$).}
\label{fig:crazymodel1}
\end{center}
\end{figure}

Analogously to the case of the reconstruction of $V(z)$, since the dependence of $H(z)$ on $w(z)$ is through an integral, for quantities that depend on  integrals of $1/H(z)$ such as $d_A(z)$ or $d_L(z)$, the information on $w(z)$ is even more diluted. This explains the weak constraints found in Fig.~\ref{fig:eos} for all the datasets except the two BAO surveys. The deterioration of the bounds comes mainly through a  degeneracy between $w_0$ and $w_1$ that can span down to $w_0\sim-100$ and $w_1\sim-100$ if only information on $d_A$ or $d_L$ is available. This degeneracy is solved with information on $H(z)$. As an example let us consider a model  lying within this degeneracy. The parameters for this model are: 
$H0 = 70.38$ km/s/Mpc, $\Omega_m = 0.28$,  $\Omega_k = 0.025$, $w_0 = -57.51$,  $w_1 = -82.63$, 
$w_2 = -28.31$ and its equation of state parameter is shown in Fig. \ref{fig:crazymodel1} along with the $\Lambda$CDM values $w=-1$ for comparison. From Eq.~(\ref{eq:wtoday}), we can see that  these parameters will still give $w=-3.19$ today, however for $z=2$, $w=-168$. 

\begin{figure}[t]
\vspace{-0.1cm}
\begin{center}
\begin{tabular}{cc}
\hspace{-0.55cm} \includegraphics[width=7.5cm]{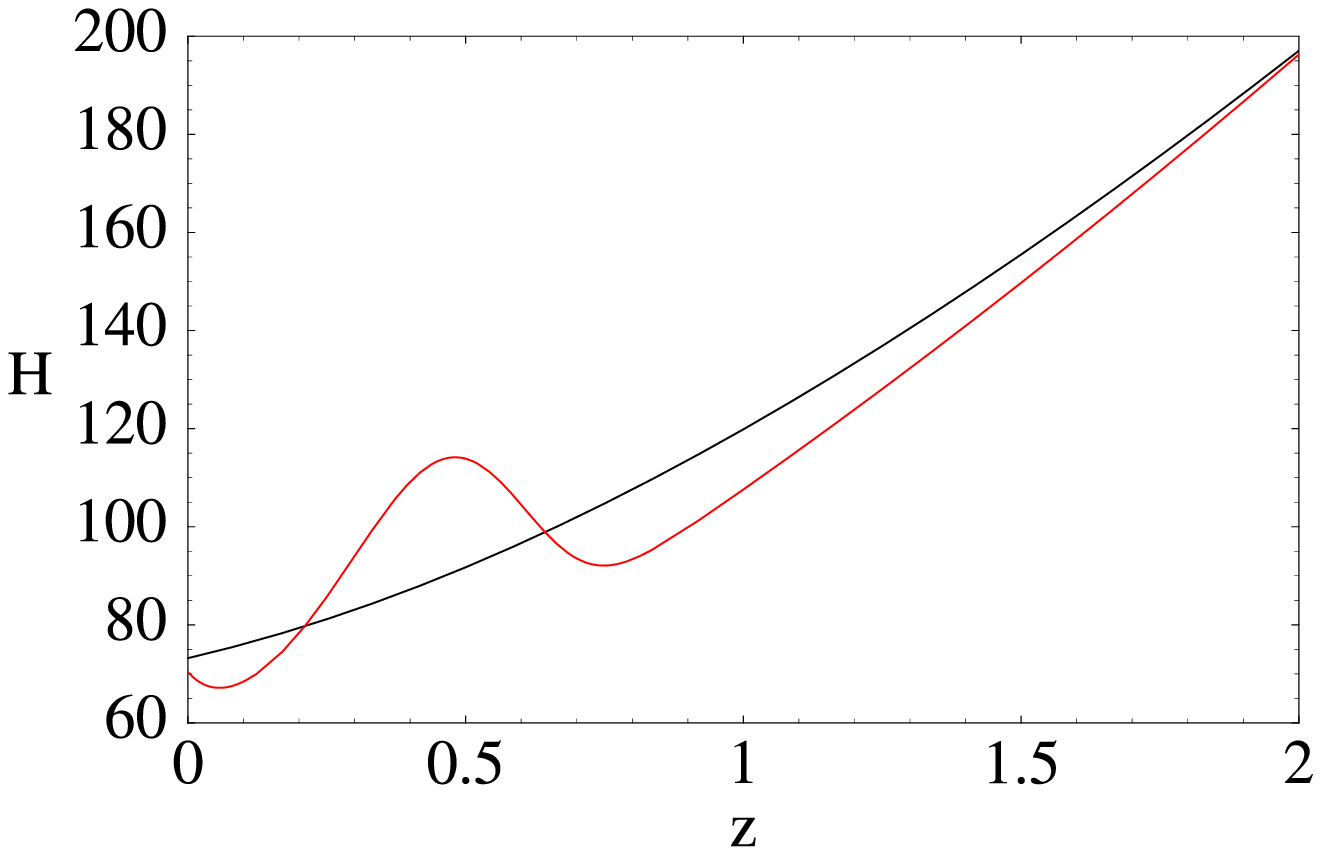} &
		 \includegraphics[width=7.5cm]{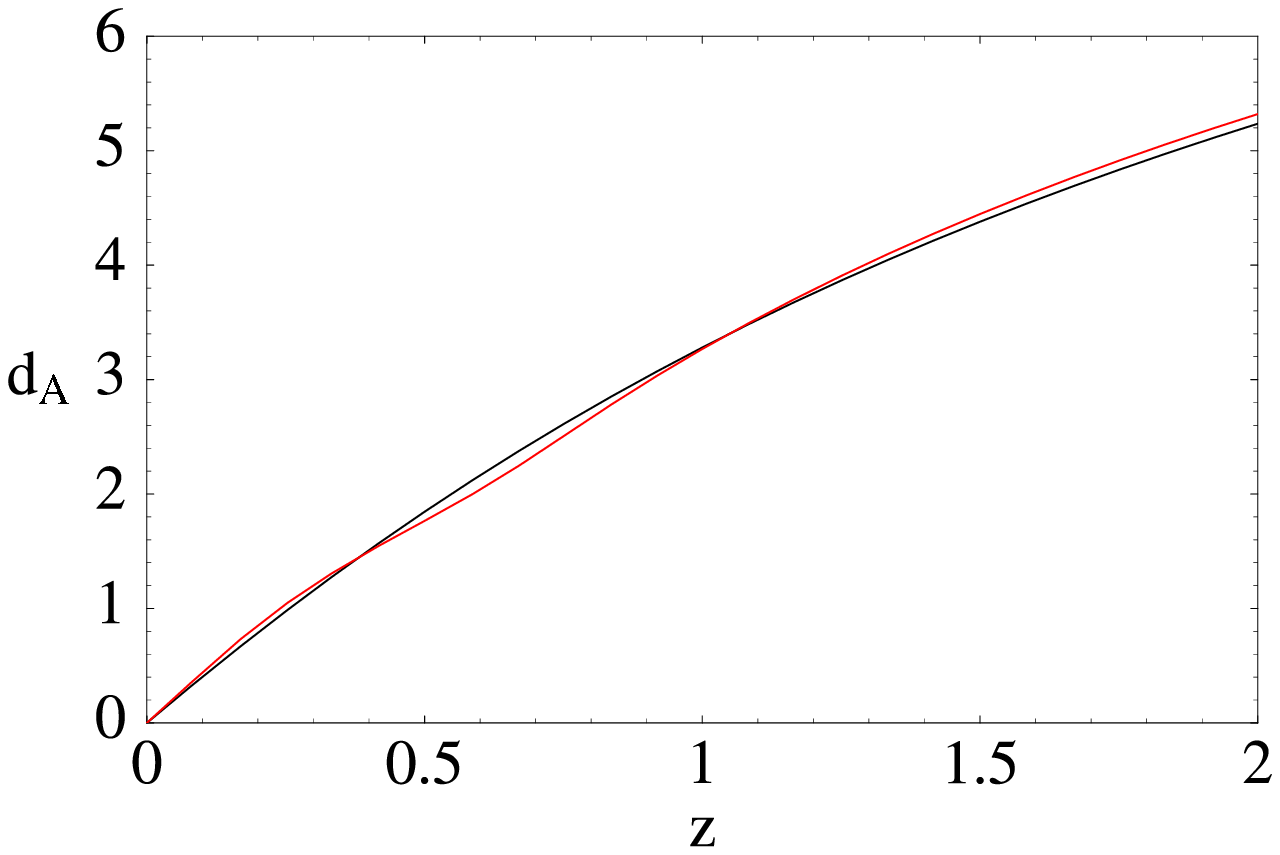} \\
\end{tabular}
\caption{\it Comparison of $H(z)$ and $d_A(z)$ for  the example model described in the text  and in Fig. \ref{fig:crazymodel1} and for the $\Lambda$CDM model.}
\label{fig:crazymodel2}
\end{center}
\end{figure}

In Fig.~\ref{fig:crazymodel2} we show the comparison between $H(z)$ and $d_A(z)$ for this model and for the $\Lambda$CDM model. From the figure it is clear that information on $d_A(z)$ alone does not suffice to discriminate between the two, while the differences in $H(z)$ between the two models are large. Even with these extreme values of the parameters, this model mimics the $d_A(z)$ behaviour of the $\Lambda$CDM, but has a  significantly different $H(z)$ which  oscillates around the $\Lambda$CDM $H(z)$. Thus, upon integrating $H(z)$ to obtain $d_A(z)$, the regions where the model is above the $\Lambda$CDM compensate the ones where it is below. A measurement of $H(z)$, however, can easily  distinguish the two models. We were not able to reproduce this behaviour to the same degree with a two-parameter description of the dark energy equation of state dynamics.
This example highlights an important open issue in dark energy studies: constraints on dark energy parameters coming from measurement of integrated quantities depend crucially on the choice of  the dark energy parameterization \cite{Maoretal01}: in the absence of a theoretical motivation for a parameterization of dark energy properties, forecasts and constraints become crucially model-dependent.

We have checked that fixing $\omega_2=0$ and $\Omega_k=0$, with just a two parameter description of the dynamics of the dark energy equation of state via $\omega_0$ and $\omega_1$, we are able to recover similar constraints to those found with alternative descriptions with two parameters.
Note that the linear parameterization in \cite{HutererTurner01, WellerAlbrech02}, corresponds to $\omega_i = 0$ for $i > 1$, and in particular $w_0= \omega_0-\omega_1$ and $w'= 2\omega_1/z_{max}$. 
Finally the linear parameterization in a \cite{ChevallierPolarski01,Linder03} , $w = w_0 + w_a z /(1 + z )$ for $|w_a|<<w_0$  can be closely approximated by $\omega_i = 0$ for $i > 2$, with the constraint (\ref{eq:wtoday}). 
Ref.~\cite{bassettCorasanitiKunz04} pointed out that a simple, 2-parameter fit may introduce biases: the 
expansion (\ref{eq:w_cheby}) allows one to include more parameters by increasing N as the observational data improve. 

The analysis shown here shows that, increasing the number of parameters seriously spoils our ability to constrain $w(z)$ except at small redshifts. As for the constraints on the potential, the stronger constraints at low redshifts are related to very pronounced degeneracies between $w_0$ and $w_1$. This degeneracies are much less important for the two BAO surveys, but the rest of the datasets considered  can  only effectively constrain the dark energy equation of state at small values of z. For a constant equation of state  ($\omega_i=0$ for $i>0$) the bound derived will roughly correspond to the narrowest allowed region at small redshifts. 

\section{Conclusions}
\label{sec:conclusions}

We have generalized to non-flat geometries the formalism of \cite{Simonetal05} to reconstruct the dark energy potential. This approach makes use of quantities similar to the horizon flow parameters used to reconstruct the inflation potential \cite{schwarz01, Leach02}. The method can, in principle, be made non-parametric, but present and forthcoming data do not  allow a fully non-parametric reconstruction.
We have therefore considered a parametric description in term of Chebyshev  polynomials which, for all our applications, we have truncated to  second order. For completeness we have also considered a reconstruction of the dark energy equation of state redshift dependence in terms of Chebyshev  polynomials, also generalizing to non-flat geometries the results of \cite{Simonetal05}.

 We have considered present  measurements of $H(z)$ from ages of passively evolving galaxies \cite{ages03,Simonetal05},  future Baryon Acoustic Oscillation (BAO) surveys and present and future type IA supernova surveys and investigated their constraints on dark energy properties.
 
 We present present and forecast constraints both on $V(z)$ (Fig.~\ref{fig:potential}) and, more interestingly, on $V(\phi)$ (Fig.~\ref{fig:potentialphi}), in sec. 4. Model building  for dark energy which  rely on simple single-field models and  provide physically motivated potentials  should satisfy the constraints shown in the  left top and middle panels of Fig.~\ref{fig:potentialphi}. In the future, the expected constraints can be as tight as those shown in the two bottom panels of Fig~\ref{fig:potentialphi}. More complicated models (multi fields etc.) should  produce a redshift evolution of the effective dark energy potential   which  satisfies the  constraints in the left top and middle panel of  Fig.~\ref{fig:potential}. The expected future constrains can be as tight as shown in the bottom panels of Fig.~\ref{fig:potential}.

 We find that  relaxing the flatness assumption increases slightly the errors on the reconstructed dark energy evolution, but does not generate significant degeneracies, provided that a modest prior on geometry is imposed $\sigma_k=0.03$ (e.g., Fig.~\ref{fig:omegak}).  
 
 Dark energy properties are best constrained at $z\lap 0.3$: this is the result of the late-time dominance of dark energy.
 Under the assumptions made here, the most crucial being the assumption of a smooth $V(z)$ or $w(z)$, we find that high redshift ($z<2$) measurements of both $H(z)$ and $d_A$ are more powerful than low $z$ measurements.  

 When constraining the redshift evolution of both the dark energy potential $V(z)$  or the dark energy  equation of state parameter $w(z)$  with measurements of integrated quantities such as $d_A$ or $d_L$, there are large degeneracies among the parameters. These degeneracies are greatly reduced or  removed with measurements of $H(z)$, such as those provided e.g., by future BAO surveys. This is illustrated in Figs.~\ref{fig:Hvsda_ground} and \ref{fig:Hvsda_space} for the potential reconstruction.   While the $H(z)$ constraint is generally weaker than the $d_A(z)$ constraints, $H(z)$ is more directly related to the dark energy properties and thus offers more powerful dark energy constraints.
We have illustrated this with an example of a model which lies on the ``$d_A$-degeneracy" for the reconstruction of $w(z)$. This model  produces $d_A(z)$ and $d_L(z)$ virtually indistinguishable from that of the $\Lambda$CDM, however the $H(z)$ are different and easily distinguishable from BAO measurements with $H(z)$ information. 

This highlights an important open issue in dark energy studies: constraints on dark energy parameters coming from measurement of integrated quantities such as $d_A$ or $d_L$ depend crucially on the choice of  the dark energy parameterization [e.g., \cite{Maoretal01}]: in the absence of a theoretical motivation for a parameterization of dark energy properties, forecasts and constraints become crucially model-dependent. The dependence of the constraints on the assumed dark energy parameterization becomes evident only when considering non-trivial deviations from a $\Lambda$CDM model (e.g., deviations from a constant $w$  or generic shape of the potential).
This issue is greatly alleviated  by measurements that carry information on $H(z)$.

\section*{Acknowledgments}
EFM acknowledges financial support from the Max Planck Institut f\"ur Physik.
LV acknowledges support of FP7-PEOPLE-2007-4-3-IRGn202182 
and CSIC I3 grant 200750I034. 
This work was supported in part by the Spanish Ministry of Education and Science (MEC) through the Consolider Ingenio-2010 program, under project CSD2007-00060 ÒPhysics of the Accelerating Universe (PAU).Ó 

\section*{References}
\bibliographystyle{JHEP}
% \bibliography{./general}

\providecommand{\href}[2]{#2}\begingroup\raggedright\endgroup

\end{document}